\documentclass[aps,prd,longbibliography,reprint,twocolumn,amsmath,amssymb,amsfonts,showpacs,footnote,superscriptaddress]{revtex4-1}

\usepackage{graphicx, epsfig, amssymb}
\usepackage{amsmath, amsfonts}
\usepackage{bm}
\usepackage{enumitem}
\usepackage[usenames]{color}
\definecolor{navyblue}{rgb}{0.0, 0.0, 0.5}
\usepackage{hyperref}
\hypersetup
	{
		colorlinks,%
		citecolor=blue,%
		linkcolor=red,%
		urlcolor=blue,%
	}

\usepackage[caption=false]{subfig}
\usepackage[utf8]{inputenc}
\usepackage{tensor}
\usepackage{soul}
\usepackage{pict2e}
\usepackage{diagbox}
\usepackage{empheq}

\usepackage{float}
\usepackage{comment}
\usepackage{array}
\usepackage{units}
\usepackage{wasysym}
\usepackage{xcolor}
\usepackage{soul}

\DeclareMathAlphabet{\mathpzc}{OT1}{pzc}{m}{it}

\newcommand{\nn}{\nonumber}

\newcommand{\phiout}{\phi^{\text{out}}_{\omega, \lambda-1/2}}
\newcommand{\phireg}{\phi^{}_{\omega, \lambda-1/2}}

\usepackage{footnote}
\usepackage{tablefootnote}

\usepackage{threeparttable}

\begin{document}

\title{Scattering from compact objects: \\ Regge poles and the complex angular momentum method}

\author{Mohamed \surname{Ould~El~Hadj}}\email{m.ouldelhadj@sheffield.ac.uk}

\affiliation{Equipe Physique
Th\'eorique, SPE, UMR 6134 du CNRS
et de l'Universit\'e de Corse,\\
Universit\'e de Corse, Facult\'e des Sciences, BP 52, F-20250 Corte,
France}

\affiliation{Consortium for Fundamental Physics,  School of Mathematics and Statistics,
University of Sheffield, Hicks Building, Hounsfield Road, Sheffield S3 7RH, United Kingdom \looseness=-1}
\author{Tom Stratton}\email{tstratton1@sheffield.ac.uk}
\affiliation{Consortium for Fundamental Physics,  School of Mathematics and Statistics,
University of Sheffield, Hicks Building, Hounsfield Road, Sheffield S3 7RH, United Kingdom \looseness=-1}
\author{Sam R. Dolan}\email{s.dolan@sheffield.ac.uk}
\affiliation{Consortium for Fundamental Physics,  School of Mathematics and Statistics,
University of Sheffield, Hicks Building, Hounsfield Road, Sheffield S3 7RH, United Kingdom \looseness=-1}
\begin{abstract}
We calculate the Regge poles of the scattering matrix for a gravitating compact body, for scalar fields and for gravitational waves in the axial sector. For a neutron-starlike body, the spectrum exhibits two distinct branches of poles, labeled surface waves and broad resonances; for ultracompact objects, the spectrum also includes a finite number of narrow resonances. We show, via a WKB analysis, that the discontinuity of the effective potential at the body's surface determines the imaginary component of the broad-resonance poles.

Next, we examine the role of Regge poles in the time-independent scattering of monochromatic planar waves. We apply complex angular momentum techniques to re-sum the partial wave series for the scattering amplitude, expressing it as a residue series evaluated at poles in the first quadrant, accompanied by a background integral. We compute the scattering cross section at several frequencies, and show precise agreement with the partial-wave calculations. Finally, we show that compact bodies naturally give rise to orbiting, glory, and rainbow-scattering interference effects.
\end{abstract}

\date{\today}

\maketitle

\tableofcontents

\section{Introduction}

The time-independent scattering of planar waves in the gravitational field of a compact body has been studied in some detail since the 1960s \cite{Hildreth1964PhDT64, Matzner:1968, Vishveshwara:1970}. A substantial literature has accumulated on \emph{black hole scattering}, focusing on the canonical scenario of a planar wave of circular frequency $\omega$ and spin $s$ \cite{Chrzanowski:1976jb} impinging upon a black hole of mass $M$ in vacuum \cite{Hildreth1964PhDT64, Matzner:1968, Vishveshwara:1970, Mashhoon:1973zz,Fabbri:1975,Sanchez:1977vz,MatznerRyan1978,Handler:1980un,Matzner:1985rjn,Futterman:1988ni,Andersson:1995vi,Glampedakis:2001cx,Dolan:2006vj,Dolan:2007ut,Dolan:2008kf,Crispino:2009xt,Cotaescu:2014jca,Gussmann:2016mkp}. A dimensionless parameter
$
M \omega = \pi r_g / \lambda
$
encapsulates the ratio of the gravitational radius $r_g = 2GM/c^2$ to the wavelength $\lambda$; herein we adopt geometric units such that $G=c=1$. The long-wavelength ($M \omega \ll 1$), short-wavelength ($M \omega \gg 1$) and intermediate regimes have been studied with a combination of perturbative \cite{DeLogi:1977dp,Dolan:2007ut,Guadagnini:2008ha,Sorge:2015yoa}, semiclassical \cite{Matzner:1985rjn, Anninos:1992ih} and numerical methods. The $s = 0$ (scalar) \cite{Matzner:1968,Sanchez:1977vz,Andersson:1995vi,Glampedakis:2001cx,Leite:2019eis}, $s=1/2$ (fermion) \cite{Dolan:2006vj,Cotaescu:2014jca}, $s=1$ (electromagnetic) \cite{Fabbri:1975, Crispino:2009xt, Crispino:2015gua} and $s=2$ (gravitational) cases \cite{MatznerRyan1978,Handler:1980un,Dolan:2008kf} have all been addressed.

Time-independent scattering by a compact body of radius $R$ with a regular center, such as a neutron star or white dwarf, has received less attention, by comparison. In such a scenario, an electromagnetic wave will not penetrate far inside the compact body; but on the other hand, a gravitational wave will pass through the body without impediment from the matter distribution. A neutron star is also expected to be effectively transparent to neutrinos. Even in such cases where the coupling to matter is negligible, the incident wave will nevertheless be scattered by the spacetime curvature. Thus, the resulting scattering pattern depends not just on $M\omega$, but also on the internal structure of the body, and on its inverse compactness or \emph{tenuity}, $R/M$. Typical values of the tenuity are: $R/M \sim 6$ for neutron stars, $\sim 1.4-9.4 \times 10^3$ for white dwarfs, $4.7 \times 10^5$ for the Sun, and $1.4 \times 10^9$ for Earth. More speculatively, we shall also consider here the case of (hypothetical) \emph{ultracompact objects} (UCOs), whose tenuity is bounded from below by the Buchdahl bound of $R / M = 9/4$.

Studies of black hole scattering typically involve the calculation of scattering amplitude(s) via partial-wave expansions.  A powerful approach for resumming these partial-wave expansions is provided by so-called complex angular momentum (CAM) techniques. In the CAM approach, the sum over partial waves is replaced by an integral, and the contour of integration is deformed into the complex-angular-momentum plane such that one collects a sum of residues of simple poles: these are the so-called \emph{Regge poles}.


The CAM theory was originally developed to deal theoretically with the propagation and diffraction of radio waves around the Earth, by Watson~\cite{Watson18} (see also the work of Sommerfeld~\cite{Sommerfeld49}). It has since been extensively used in several domains of physics involving resonant scattering theory (see, e.g., Refs.~\cite{deAlfaro:1965zz,Newton:1982qc,Watson18,Sommerfeld49,Nussenzveig:2006,Grandy2000,Uberall1992,AkiRichards2002,AkiRichards2002,Gribov69,
Collins77,BaronePredazzi2002,DonnachieETAL2005} as well as references therein for various applications in quantum mechanics, nuclear physics, electromagnetism, optics, seismology and high energy physics). Since the pioneering work in 1994 by Andersson and Thylwe~\cite{Andersson:1994rk,Andersson:1994rm} the CAM theory has been successfully applied to black hole scattering scenarios. They described the scattering of scalar waves by a Schwarzschild black hole using CAM techniques, and showed that, in this case, the Regge poles are associated with ``surface waves'' localised near the unstable light ring at $ r = 3M $ (\textit{i.e.}, the so-called photon sphere). Later, Decanini, Folacci and Jensen showed that the complex frequencies of weakly damped quasinormal modes (QNMs) are Breit-Wigner resonances generated by the surface waves previously mentioned and, using the concept of Regge trajectories, they were able to construct semiclassically the spectrum of the QNM complex frequencies~\cite{Decanini:2002ha}. Recently, Folacci and Ould~El~Hadj have shown that CAM machinery can be used for precise numerical calculations of scattering cross sections of scalar fields, electromagnetic fields~\cite{Folacci:2019cmc} and gravitational waves~\cite{Folacci:2019vtt} on a Schwarzschild space-time. Moreover, using the third WKB approximation to obtain the asymptotic expressions for the lowest Regge poles~\cite{Decanini:2009mu} and associated residues, they have been able to provide an analytical approximation describing accurately both the ``glory'' and a large part of the ``orbiting'' oscillations in black hole scattering cross sections in the short-wavelength regime. An accurate approximation for the black hole absorption cross section has also been derived via an analysis of Regge poles~\cite{Decanini:2011xi,Decanini:2011xw}. In addition, an alternative description of gravitational radiation from black holes based on CAM theory was developed in Ref.~\cite{Folacci:2018sef}.

CAM techniques are yet to be fully applied to compact-body scattering, it appears, with one notable exception: in 1991, Chandrasekhar and Ferrari \cite{ChandrasekharIV:1992ey} examined the Regge pole spectra for relativistic stellar models.
Their study focussed on resonant modes (poles with small imaginary parts corresponding to trapped $w$-modes or fluid modes), and it employed CAM theory to calculate the flow of gravitational energy through a star. 

There is a close relationship between Regge poles and QNM frequencies; both are sets of poles of the scattering matrix, with the latter lying in the complex-frequency plane. The excitation of black hole QNMs leads to distinctive signatures in \emph{time-dependent} scattering scenarios. For example, in the immediate aftermath of a black hole merger, the perturbed black hole returns to a quiescent state via a ``ringdown phase'' in which its gravitational-wave signal is well described as a sum of quasinormal modes \cite{Giesler:2019uxc}. The QNM spectrum of compact bodies have been well studied by several authors  \cite{Detweiler:1985zz, Kokkotas:1986gd, Chandrasekhar449, Kokkotas:1992ka, Leins:1993zz, Andersson:1995ez, Andersson1996}, and are reviewed in Ref.~\cite{Kokkotas:1999bd}.


Time-independent scattering by compact bodies generates (in principle) a \emph{rainbow scattering} phenomenon in the short wavelength regime ($M \omega \gg 1$)~\cite{Dolan:2017rtj, Stratton:2019deq}. The rainbow scattering phenomenon is linked to the formation of a cusp caustic in the incident wavefront (see Fig.~1 in Ref.~\cite{Stratton:2019deq}), and to the stationary point in the geodesic deflection function associated with a ray (that is, a null geodesic) that passes somewhat inside the body. For less-dense bodies, such as stars and white dwarfs, the cusp caustic forms at a distance $d \sim R^2 / (4M)$ downstream of the body, and the rainbow angle is $\theta_r \sim 4M/R$ (e.g.~for the Sun, $d \approx 550 \, \text{a.u.}$ and $\theta_r \approx 1.8 \, \text{arcsec}$) \footnote{The rainbow angle $\theta_r \approx 4M/R$ is distinct from the Einstein ring angle of $\theta_E \approx \sqrt{4M/r}$}. For compact bodies such as neutron stars, however, the cusp caustic forms near the surface of the body, and the rainbow angle is large ($\theta_r \gtrsim 59.6^\circ$ for $R/M = 6$; see Table I in Ref.~\cite{Stratton:2019deq}). In this case, the rainbow angle is -- in principle at least -- a diagnostic of the matter distribution within the body and thus its nuclear equation of state.

Other recent works on the theme of scattering by a compact body include studies of: scattering by compact objects with an absorbing surface \cite{Nambu:2019sqn}; time-domain simulations with finite-element-method methods \cite{He:2019orl}; gravitationally-induced interference patterns in flavour oscillations in neutrino astronomy \cite{Alexandre:2018crg}; and the use of continuous sources of gravitational waves to probe stellar structure \cite{Marchant:2019swq}.

The remainder of this paper is organised as follows. Sec.~\ref{SecII} reviews the theory of scalar waves on a spherically symmetric spacetime of a compact object. Here we describe our model spacetime (\ref{SecIIa}), the effective potentials for wave scattering (\ref{subsec:Potentials}) and the physical boundary conditions that define the $S$-matrix (\ref{subsec:bc}). Sec.~\ref{sec:RP} focusses on the Regge pole spectrum. Here we outline the link between Regge poles and QNMs (\ref{subsec:QNMs_RPs}), we describe the numerical method (\ref{subsec:method}), and we present new numerical results for the spectrum (\ref{subsec:results1}) which exhibits three distinct branches of Regge poles (see e.g.~Fig.~\ref{RP_R226_approx_2Mw_3_6_s_1}). In Sec.~\ref{subsec:WKB} we apply the WKB method to obtain an approximate formula for the ``broad resonance'' branch. Sec.~\ref{sec:CAM} concerns the application of CAM techniques to the calculation of the scattering cross section. Here we review the standard partial-wave expansion of the scattering amplitude (\ref{subsec:partial}), we apply CAM techniques to write this as the sum of a residue series and a background integral (\ref{SecIIc}), and we present a selection of numerical results for $d\sigma/d\Omega$ (\ref{subsec:results2}). We conclude with a discussion in Sec.~\ref{sec:conclusions}. In an appendix, we show which quadrants of the complex $\lambda$-plan contain Regge poles of the $S$-matrix.

\section{Waves on a compact-body spacetime} \label{SecII}



\subsection{The model}
\label{SecIIa}

The gravitating source is assumed to be spherically-symmetric, such that in a coordinate system $\{t,r,\theta,\varphi\}$, the object is described by a diagonal metric $g_{\mu \nu}$ and the line element
\begin{equation}\label{Line_elem}
 ds^2 = g_{\mu \nu} dx^\mu dx^\nu = -f(r) dt^2+h(r)^{-1}dr^2+r^2d\sigma_2^2
\end{equation}
where $d\sigma_2^2 = d\theta^2 + \sin^2 \theta d\varphi^2$ denotes the metric on the unit 2-sphere $S^2$. In the vacuum exterior of the star ($r>R$), the radial functions $f(r)$ and $h(r)$ depend only on $M$, the total mass of body: $f(r)=h(r)=1-2M/r$ by Birkhoff's theorem~\cite{VojeJohansen:2005nd}. In the interior, $f(r)$ and $h(r)$ depend on the matter distribution and equation of state (EoS).

A widely-studied model is that of a polytropic star, with an EoS $p(\rho) = \kappa \rho^{1+1/\hat{n}}$, where $\hat{n}$ is the polytropic index (see e.g.~\cite{Stratton:2019deq}). Here we shall consider a special case: an incompressible perfect fluid ball of uniform density described by Schwarzschild's interior solution for an incompressible fluid \cite{Shapiro1983}, with
\begin{subequations}
\begin{align}
\rho &= \frac{M}{\frac{4}{3} \pi R^3} , \\
p &= \rho \frac{\beta(R) - \beta(r)}{ \beta(r) - 3 \beta(R)} , \\
\beta(x) &= \sqrt{3 - 8 \pi \rho x^2},
\end{align}
\end{subequations}
and metric functions
\begin{subequations}\label{Interior_Solution}
\begin{align}\label{Interior_Solution_f}
 f(r) &= \frac{1}{4}\left(1-\frac{2 M r^2}{R^3}\right)+\frac{9}{4}\left(1-\frac{2M}{R}\right) \nn \\
       & \quad -\frac{3}{2} \sqrt{\left(1-\frac{2M}{R}\right)\left(1-\frac{2 M r^2}{R^3}\right)}, \\
 h(r) &= 1-\frac{2 M r^2}{R^3} . \label{Interior_Solution_h}
\end{align}
\end{subequations}
The constant-density model can be thought of as representing the $\hat{n} \rightarrow 0$ limit of the family of polytropes.
The radial function $h(r)$ is $C^0$ at the surface of the star $r=R$ (i.e.~continuous but not differentiable), and the radial function $f(r)$ is $C^1$ there (i.e.~once-differentiable).
For a general polytrope with $\hat{n}>0$ and $m$ the smallest integer such that $m \geq \hat{n}$, $h(r)$ is $C^m$ and $f(r)$ is $C^{m+1}$ at the surface.
As we shall see, the breakdown of smoothness leads to consequences for the Regge pole spectrum.

We shall consider a scalar wave $\Phi(x)$ propagating on the compact body spacetime, governed by the Klein-Gordon equation
\begin{equation}
\Box \Phi \equiv \frac{1}{\sqrt{-g}} \partial_{\mu} \left( \sqrt{-g} g^{\mu \nu} \partial_{\nu} \Phi \right) = 0
\end{equation}
where $g^{\mu \nu}$ is the inverse metric and
$
g 
$
 is the metric determinant. Performing a standard separation of variables,
\begin{equation}
\Phi = \frac{1}{r} \sum_{\omega \ell m} \phi_{\omega \ell}(r)  Y_{lm}(\theta, \phi) e^{-i \omega t} ,
\label{eq:sepvariables}
\end{equation}
leads to a radial equation of the form
\begin{equation}
\label{H_Radial_equation}
\left[\frac{d^{2}}{dr_{\ast}^{2}}+\omega^{2}-V_{\ell}(r)\right]\phi_{\omega\ell}= 0,
\end{equation}
where $V_{\ell}(r)$ is the effective potential, and $r_\ast$ denotes the \emph{tortoise coordinate} defined by
\begin{equation}
\frac{dr}{dr_\ast} =\sqrt{f(r)h(r)} .  \label{eq:tortoise}
\end{equation}

\subsection{Effective potentials}\label{subsec:Potentials}
The effective potential for the scalar field in Eq.~(\ref{H_Radial_equation}) is $V_{\ell}(r) = V_{\ell}^{(s=0)}(r)$, where we define
\begin{equation}\label{Inside_Potentiel}
  V_{\ell}^{(s)}(r) \equiv f(r)\left[\frac{\ell(\ell+1)}{r^2}+\frac{\beta_s \, h(r)}{2r}\left(\frac{f'(r)}{f(r)}+\frac{h'(r)}{h(r)}\right)\right] ,
\end{equation}
where $\beta_s \equiv 1-s^2$.
Remarkably, the radial equation for axial gravitational perturbations is identical to Eq.~(\ref{H_Radial_equation}) but with an effective potential $V_{\ell}^{\text{ax}}(r)$ where \cite{Cardoso:2014sna}
\begin{eqnarray}\label{RW_Potentiel_axialGW}
V_{\ell}^{\text{ax}}(r) = V_{\ell}^{(s=2)} + 8 \pi f(r) (p - \rho)  .
\end{eqnarray}
Outside the star in the vacuum region ($r>R$), the effective potentials reduce to the Regge-Wheeler potential,
\begin{equation}\label{RW_Potentiel}
V^{(s)}_{\ell}(r) =\left(1-\frac{2M}{r}\right)\left(\frac{\ell(\ell+1)}{r^2}+\frac{2M \beta_s}{r^3}\right)
\end{equation}
with $s=0$ in the scalar-field case, and $s=2$ in the axial gravitational-wave case. In the exterior, the tortoise coordinate $r_\ast$ reduces to $r_\ast = r+ 2M \ln [r/(2M) -1]+k$, where $k$ is a constant that is chosen such that $r_\ast(r=0) = 0$ and $r_\ast(r)$ is a continuous function.

\begin{figure}[htb]  
\centering
 \includegraphics[scale=0.50]{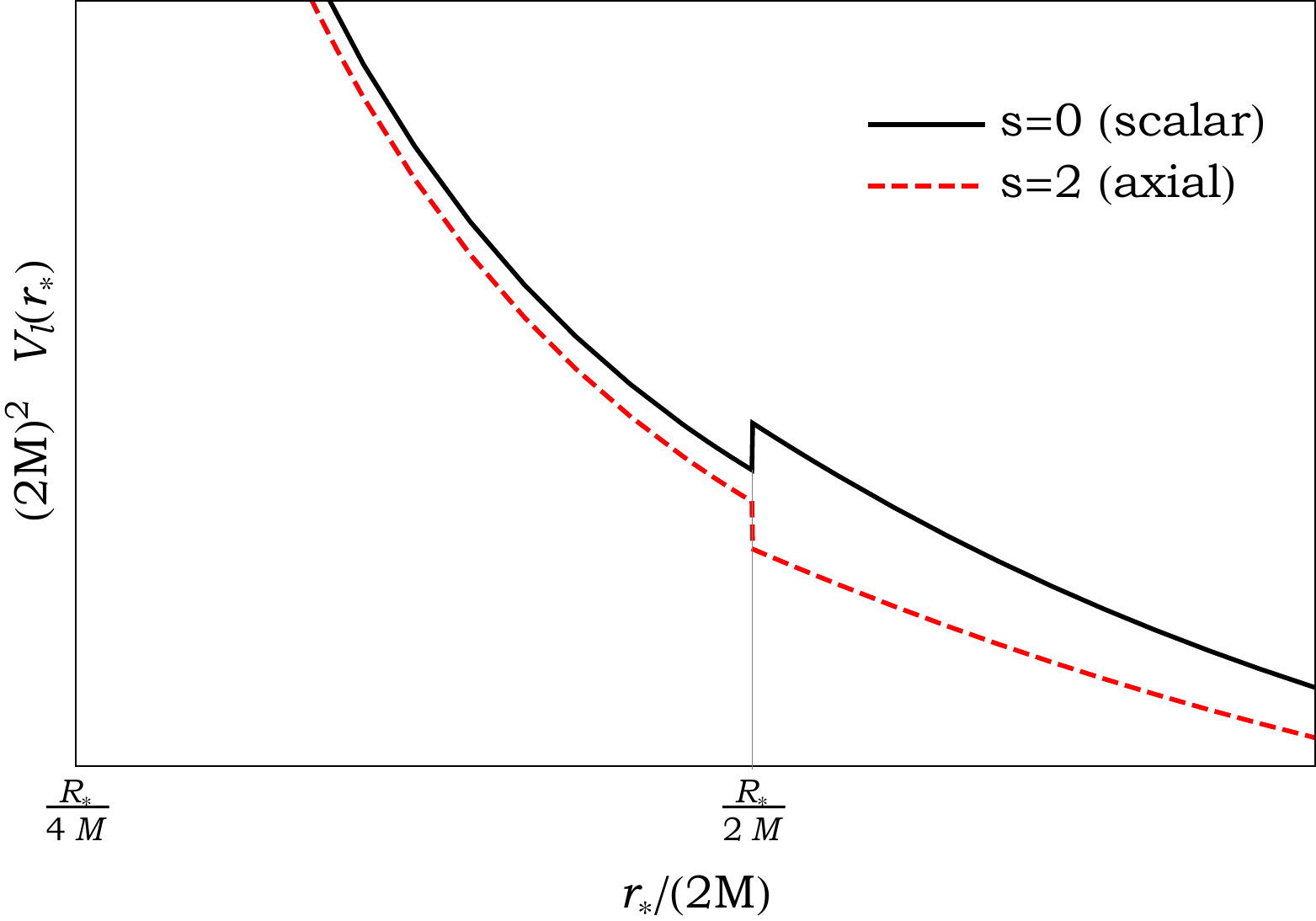} \\ \vspace{0.2cm}
 \includegraphics[scale=0.50]{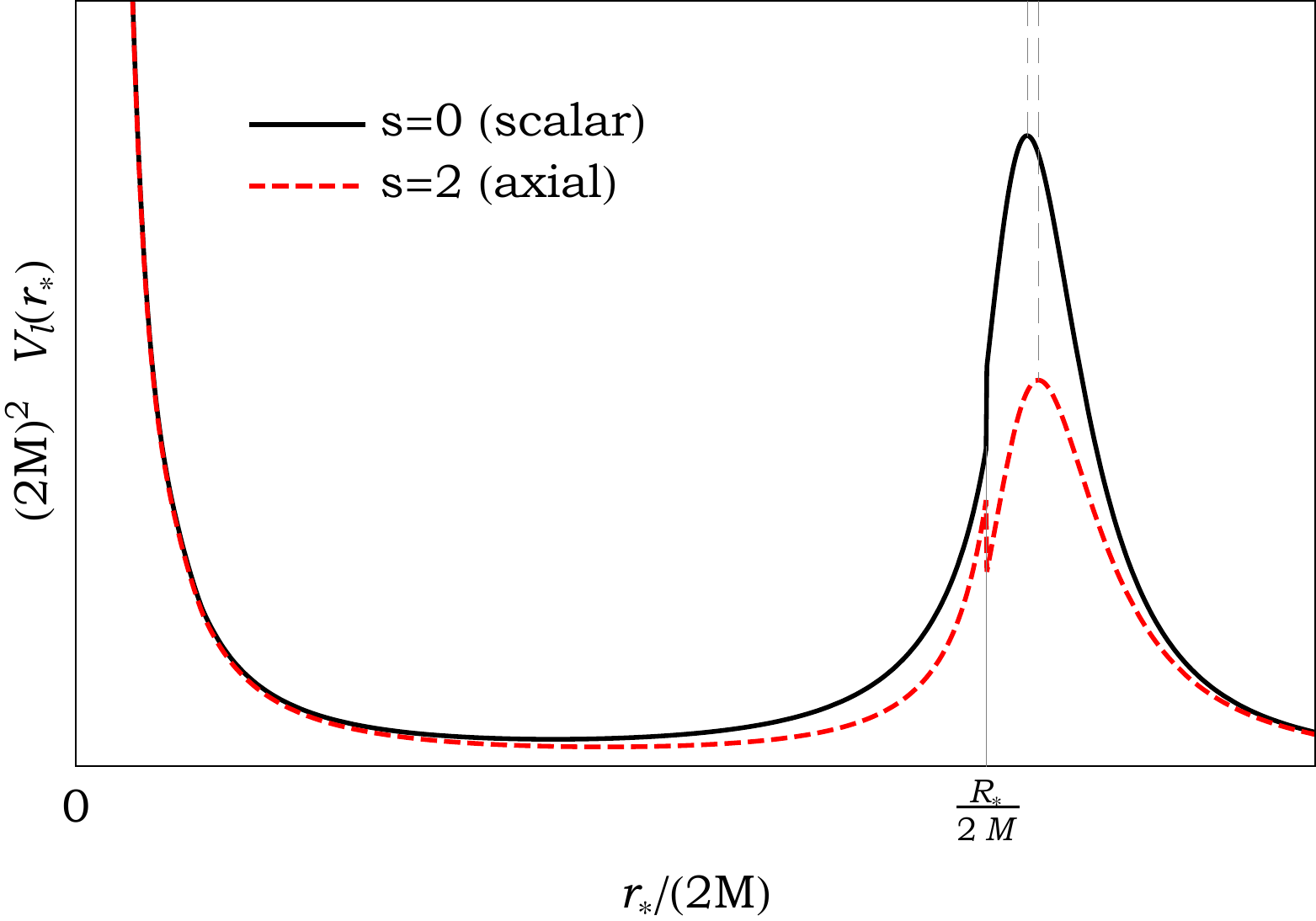}
\caption{\label{RP_approx_2Mw_3_6_s_1} The effective potential $V_\ell$ for a quadrupole ($\ell =2$) perturbation of a compact body of constant density and tenuity $R/M = 6$ (upper) and $R /M = 2.26$ (lower). The scalar field potential - Eq.~(\ref{Inside_Potentiel}) - and axial gravitational-wave potential - Eq.~(\ref{RW_Potentiel_axialGW}) - are indicated as solid/black lines and dotted/red lines, respectively. The horizontal axis is the tortoise coordinate $r_\ast/(2M)$ defined in Eq.~(\ref{eq:tortoise}).}
\label{fig:Veff}
\end{figure}

Effective potentials for the incompressible model are shown in Fig.~\ref{fig:Veff}, for two cases: (i) a neutron-star model with $R = 6M$, and (ii) a ultracompact object (UCO \cite{Macedo:2018yoi}) with $R = 2.26M$. 
In both cases we observe a discontinuity in $V_{\ell}(r)$ across the star's surface, due to the discontinuity in the density $\rho$ (which implies the $C^0$ property of $h(r)$). The jump in the potential takes opposite signs in the scalar-field and gravitational-wave cases, with
\begin{subequations}
\begin{align}
\Delta V_{\ell}^{(s=0)} &= + \frac{3Mf(R)}{R^3} , \\
\Delta V_{\ell}^{\text{ax}} &= - \frac{3Mf(R)}{R^3} ,
\end{align}
\end{subequations}
where
\begin{equation}
\Delta V_{\ell} \equiv \lim_{\epsilon \rightarrow 0} \left\{ V_{\ell}(R + \epsilon) - V_{\ell}(R - \epsilon) \right\} . \label{def:DeltaV}
\end{equation}

In the UCO case ($R < 3M$), the effective potential has a maximum near the light-ring at $r=3M$, and there is a trapping region, as shown in Fig.~\ref{fig:Veff}.

\subsection{Boundary conditions and scattering}\label{subsec:bc}
The modes $\phi_{\omega \ell}$ in Eq.~(\ref{eq:sepvariables}) should have a regular behaviour at the center of the object ($r = 0$), and inspection of the radial equation (\ref{H_Radial_equation}) shows that
\begin{equation}\label{bc_1_in}
\phi_{\omega  \ell}(r) \, \scriptstyle{\underset{r \to 0}{\sim}} \,
\displaystyle{r^{\ell+1}}.
\end{equation}
At the boundary of the compact object, the potential is $C^0$ and thus the mode is $C^2$.
The asymptotic behaviour of the modes far from the body ($r \to +\infty$, or equivalently $r_\ast \to +\infty$) is
\begin{equation}\label{bc_2_in}
\phi_{\omega  \ell}(r) \scriptstyle{\underset{r_\ast \to +\infty}{\sim}}
\displaystyle{ A^{(-)}_\ell (\omega) e^{-i\omega r_\ast} + A^{(+)}_\ell (\omega) e^{+i\omega r_\ast}}.
\end{equation}
With the complex coefficients $A^{(\pm)}_\ell (\omega)$ we then define the \emph{$S$-matrix elements},
\begin{equation}\label{Matrix_S}
  S_{\ell}(\omega) =  e^{i(\ell+1)\pi} \, \frac{A_{\ell}^{(+)}(\omega)}{A_{\ell}^{(-)}(\omega)}.
\end{equation}
We now consider the poles of $S_{\ell}(\omega)$ in the complex plane.


\section{The Regge pole spectrum}\label{sec:RP}

 \subsection{Quasinormal modes and Regge poles}\label{subsec:QNMs_RPs}
 Mathematically, there is a close relationship between quasinormal modes and Regge poles; they are both sets of poles of the scattering matrix. Physically, quasinormal modes are most relevant to time-dependent scattering scenarios, and Regge poles to time-independent scattering scenarios.

The \emph{quasinormal mode spectrum} is the set of frequencies $\{ \omega_{\ell n} \}$ in the complex-$\omega$ plane at which the scattering matrix $ S_{\ell}(\omega)$ has a simple pole for an integer value of $\ell$ (so $\ell \in \mathbb{N}$ and $\omega_{\ell n} \in \mathbb{C}$).

The \emph{Regge pole spectrum} is the set of angular momenta $\lambda_{n}(\omega) \equiv \ell_{n}(\omega) + 1/2$ in the complex-$\lambda$ plane at which the scattering matrix has a simple pole for a real value of $\omega$ (so $\omega \in \mathbb{R}$ and $\lambda_{n}(\omega) \in \mathbb{C}$). Here $n$ is an index for enumerating the discrete spectrum of poles. In all cases considered in this work, the simple poles of $S_{\ell}(\omega)$ arise as simple zeros of $A_{\ell}^{(-)}(\omega)$.

The quasinormal mode spectrum of spherically symmetric compact objects has been studied in some detail in Refs.~\cite{Kokkotas:1986gd,Kokkotas:1992ka,Andersson:1997eq,Kokkotas:1999bd,Leins:1993zz}.
Newly formed neutron stars, the remnants of supernovae collapse, are predicted to pulsate with a large initial energy, and fluid pulsations will generate gravitational waves. In 1967, Thorne and Campolattaro \cite{thorne1967non} classified the fluid modes of a relativistic compact body by analogy with the fluid modes of a Newtonian body, with the addition of a damping time due to the emission of GWs. Two decades later, the subject was examined again \cite{Detweiler:1985zz,Kokkotas:1986gd}, and Kokkotas and Schutz \cite{Kokkotas:1992ka} showed the existence of an additional family of modes, dubbed $w$-modes. These modes are characterised by a negligible excitation of fluid motion, and in the axial sector, by no fluid motion at all. They are highly damped and correspond to excitations of the dynamical perturbed space-time. For a review of (gravitational) quasinormal modes in relativistic stars and black holes, see Ref.~\cite{Kokkotas:1999bd}.

The $w$-modes (quasinormal modes) may be divided into three branches:
	\begin{enumerate}
   \item Curvature modes, the standard $w$-modes extant for all relativistic stars. The less compact the star the more rapid the damping (Im$\{\omega\}$ increases with $R/M$).
   \item Interface modes ($\omega_{\text{II}}$-modes \cite{Leins:1993zz}), characterised by very rapid damping (i.e.~large negative imaginary part of $\omega_{\ell n}$). This branch of modes is somewhat similar to modes for acoustic waves scattered by a hard sphere.
   \item Trapped modes \cite{Chandrasekhar449}: These modes may exist when the effective radial potential has a cavity region, which is the case for UCOs ($R/M < 3$). 
   The number of trapped modes increases with the depth of the potential well, and the damping rate decreases.
	\end{enumerate}

We now move on to consider the Regge poles of compact bodies, which have received comparatively less attention \cite{ChandrasekharIV:1992ey}.

 \subsection{Numerical method}\label{subsec:method}


Leins, Nollert and Soffel \cite{Leins:1993zz} have developed a method for calculating the polar QNM frequencies of spherically symmetric spacetimes. Benhar, Berti and Ferarri (BBF) have extended this method to the axial sector. Here we shall present a generalisation of the BBF method for axial \emph{and} scalar QNM frequencies ($s=2$ and $s=0$ respectively). The method is equally valid, \textit{mutatis mutandis}, for finding Regge poles $\{\lambda_n(\omega)= \ell_n +1/2\}$ of $ S_{\ell}(\omega)$.  The solution of the Regge-Wheeler equation is written in a power-series form as follows:
\begin{eqnarray}\label{Power_series}
 \phi_{\omega_\ell}(r)&=&\left(\frac{r}{2M}-1\right)^{i 2M\omega} \,e^{i\omega r} \sum_{n=0}^{+\infty}a_n\left(1-\frac{b}{r}\right)^n , \nonumber \\
                      &=& e^{i\omega r_*(r)}\sum_{n=0}^{+\infty}a_n\left(1-\frac{b}{r}\right)^n .
\end{eqnarray}
where $r = b$ is some point outside the star. By substituting \eqref{Power_series} in the Regge-Wheeler equation, it can be shown that the coefficients $a_n$ satisfy a four-term recurrence relation of the form:
\begin{equation}\label{Recurrence_4_terms}
\alpha_n a_{n+1} + \beta_n a_{n} +\gamma_{n} a_{n-1} +\delta_{n} a_{n-2}  = 0, \quad \forall n\geq 2 ,
\end{equation}
where
\begin{subequations}
\begin{eqnarray}\label{Coeffs_3_termes}
 && \alpha_n = n (n+1)\left(1-\frac{2M}{b}\right),    \\
 && \beta_n  = 2i\omega b n + 3\left(\frac{2M}{b}\right)n^2-2n^2 ,  \\
 && \gamma_n = \left(1-\frac{6M}{b}\right)n(n-1)-\frac{2M\beta_s}{b}-\ell(\ell+1) ,  \\
 && \delta_n = \left(\frac{2M}{b}\right)\left(n(n-2)+\beta_s\right) .
\end{eqnarray}
\end{subequations}
To determine the initial conditions $a_0$ and $a_1$, we impose the continuity of $\phi_{\omega_\ell}(r)$ and its derivative in $r = b$, according to Eq~\eqref{Power_series}, to obtain:
\begin{eqnarray}\label{Initial_Conds}
 && a_0 = e^{i\omega r_*(b)\phi_{\omega_\ell}(b)} ,  \\
 && a_1 = b e^{-i\omega r_*(b)}\left(\frac{-i\omega b}{b-2M}\phi_{\omega_\ell}(b)+ \frac{d}{dr}\phi_{\omega_\ell}(r)\Big{|}_{r=b}\right) .
\end{eqnarray}
Here, to obtain $\phi_{\omega_\ell}(b)$ and its derivative in $r=b$, we need to solve Eq.~\eqref{H_Radial_equation} numerically inside the star, and then to extend the solution outside, up to $r=b$, by solving the Regge-Wheeler equation numerically.

It is important to note that, the \textit{continued fraction method} applies to three-term recurrence relations. Leins~\emph{et al.} and BBF use the method introduced by Leaver in
the case of Reissner-Nordst\"om black holes \cite{Leaver:1990zz} to reduce four-term recurrence relations [as in \eqref{Recurrence_4_terms}] to three-term recurrence relations by using a Gaussian elimination step.

We do not use the Gaussian elimination step. Instead we find solutions of \eqref{Recurrence_4_terms} by adapting the \textit{Hill determinant approach} employed by Majumdar and Panchapakesan~\cite{mp} to solve the black hole recurrence formula of Leaver \cite{Leaver:1985ax}. The nontrivial solutions of \eqref{Recurrence_4_terms} exist when the Hill determinant vanishes:
\begin{equation}
\label{Determinant_Hill_4_termes}
D   =  \begin{vmatrix}
   \beta_0         &  \alpha_0    &        0             &       0                    &          0                &     \ldots          &     \ldots        &    \ldots \\
   \gamma_1        &  \beta_1      &     \alpha_1     &       0                    &          0                &     \ldots          &     \ldots        &    \ldots  \\
    \delta_2       &   \gamma_2     &    \beta_2       &   \alpha_2             &         0                &     \ldots           &    \ldots         &    \ldots  \\
    \vdots         &      \ddots     &     \ddots       &    \ddots                &       \ddots          &     \ddots           &   \ldots          &    \ldots \\
     \vdots        &     \vdots      &     \delta_{n-1}      &    \gamma_{n-1}      &     \beta_{n-1}     &      \alpha_{n-1} &    \ddots         &    \ldots  \\
     \vdots        &  \vdots        &     \vdots        &     \delta_n                &    \gamma_{n}    &     \beta_{n}        &     \alpha_{n}  &    \ddots     \\
     \vdots        &  \vdots        &     \vdots        &     \vdots                &    \ddots            &     \ddots            &     \ddots       &    \ddots
                        \end{vmatrix}
       = 0.
\end{equation}
Letting $D_n$ be the determinant of the $n \times n$ submatrix of $D$ (with diagonal $\{\beta_0,\beta_1,....\beta_n\}$), by \eqref{Recurrence_4_terms}
\begin{equation}
\label{derecurrence_4_termes}
D_n=\beta_n D_{n-1} - \gamma_{n}\alpha_{n-1}D_{n-2} + \delta_n \alpha_{n-1} \alpha_{n-2} D_{n-3} ,
\end{equation}
with the initial conditions
 \begin{equation}
 \label{Determinant_initial_conds}
 \begin{split}
    D_0 &=\beta_0, \\
    D_1 &=\beta_1\beta_0-\gamma_1\alpha_0 , \\
    D_2 &=\beta_0(\beta_1 \beta_2 - \alpha_1 \gamma_2)- \alpha_0(\alpha_1 \delta_2 -\gamma_1 \beta_2) .
 \end{split}
 \end{equation}
Equivalently, 
\begin{align}
\label{recurrence_Hill_RN_4_termes}
D_n    & =\left(\prod_{k=1}^{n+1} k^2\right) P_{n+1} , \nonumber\\
       &  = 1\times 2^2 \ldots (n-2)^2(n-1)^2n^2(n+1)^2 P_{n+1},
\end{align}
where
\begin{eqnarray}
\label{recurrence_Hill_RN_4_termes_bis}
P_n = & & \left(\frac{\beta_{n-1}}{n^2}\right)P_{n-1}-\left(\frac{\gamma_{n-1}}{n^2}\right)\left(\frac{\alpha_{n-2}}{(n-1)^2}\right)P_{n-2} \nonumber \\
  &+& \left(\frac{\delta_{n-1}}{n^2}\right) \left(\frac{\alpha_{n-2}}{(n-1)^2}\right) \left(\frac{\alpha_{n-3}}{(n-2)^2}\right) P_{n-3}
,
\end{eqnarray}
with the initial conditions
 \begin{equation}
 \begin{split}
    P_1&=\beta_0 , \\
    P_2&=\frac{\beta_0\beta_1-\gamma_1\alpha_0}{4} ,\\
    P_3&=\frac{\beta_0(\beta_1 \beta_2 - \alpha_1 \gamma_2)- \alpha_0(\alpha_1 \delta_2 -\gamma_1 \beta_2)}{36} .
 \end{split}
 \end{equation}

Regge poles (QNM frequencies) are found by fixing $\omega$ ($\lambda$) and numerically finding roots $\lambda_n$ ($\omega_n$) of $D_n$, i.e.,~$D_n(\lambda_n,\omega)=0$ ($D_n(\lambda,\omega_n)=0$).

 \subsection{Numerical results: The spectrum}\label{subsec:results1}
In this section we show numerical results for the Regge pole spectrum of a compact body, in two particular cases: (i) a neutron-starlike body with tenuity $R / M = 6$, and (ii) a UCO with tenuity $R / M = 2.26$, close to the Buchdahl bound. Results for the frequencies $M \omega = 3/2$ and $M \omega = 8$ are compared, for the scalar-field ($s=0$) and axial GW ($s=2$) cases.

\begin{figure}[htb]
\centering
 \includegraphics[scale=0.50]{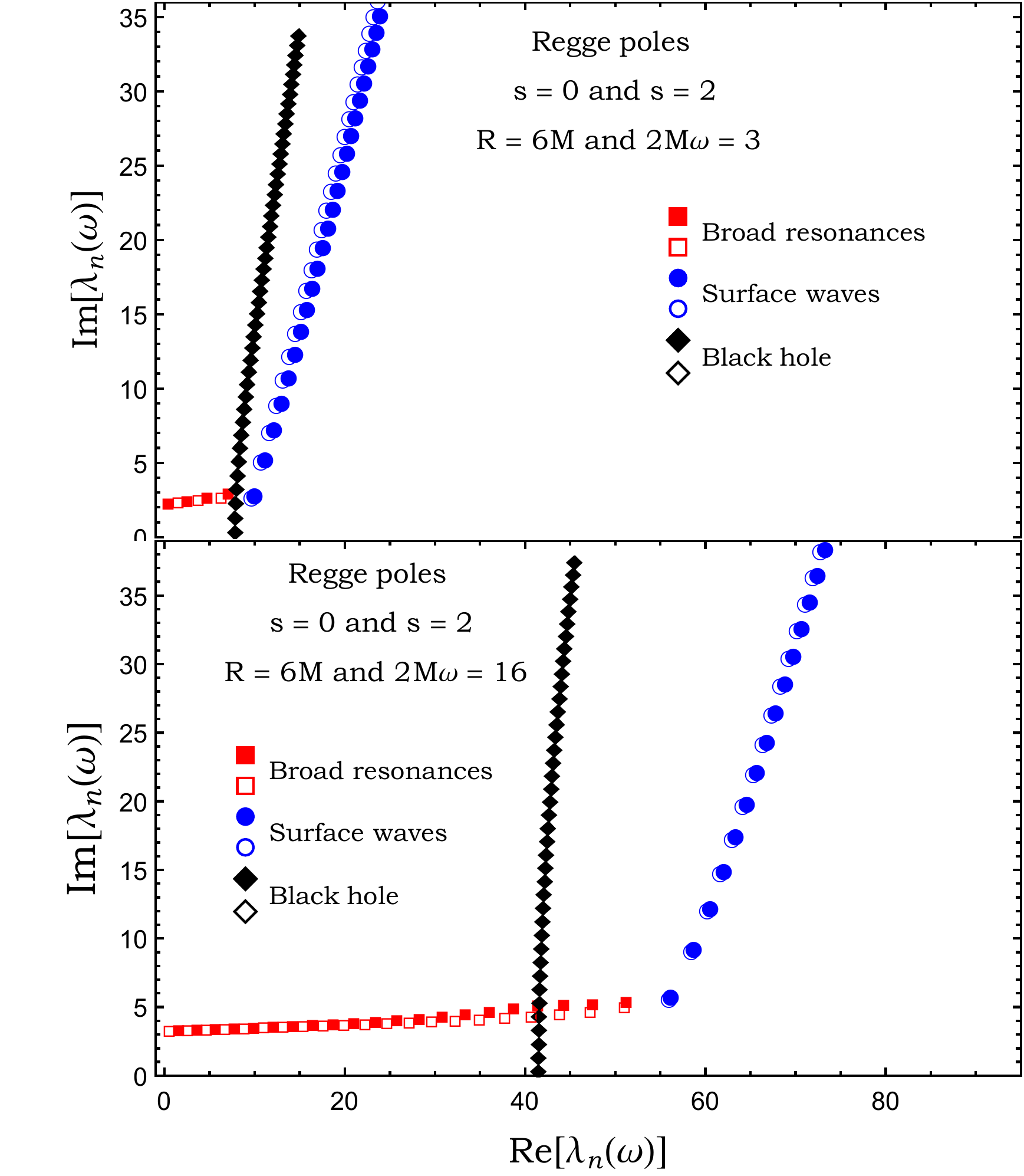}
\caption{\label{RP_R6_approx_2Mw_3_6_s_1} The Regge poles $\lambda_n(\omega)$ for the scalar field (unfilled markers) and for the axial gravitational waves (filled markers). 
}
\label{fig:rp1}
\end{figure}

Figure \ref{fig:rp1} shows the Regge pole spectrum for a neutron-starlike body ($R/M = 6$). We see that there are two branches of Regge poles in the first quadrant, which meet near $\text{Re}( \lambda_n(\omega) ) \sim \omega b_0$, where $b_0$ is the impact parameter for the null ray which grazes the surface of the compact body. The axial GW $s=2$ modes (filled markers) are typically close to their scalar-field $s=0$ counterparts (unfilled markers), as might be anticipated from the similarities in the effective potentials for the two cases (see Fig.~\ref{fig:Veff}).

In Appendix \ref{appen} we present an argument for why no Regge poles are expected to be located in the fourth quadrant of the complex-$\lambda$ plane for $\omega>0$. As a consistency check, we have scanned the fourth quadrant with the numerical method above, and have found no evidence for Regge poles.

Exploring the second and third quadrants, where $\text{Re}[{\lambda}] < 0$, requires a choice to be made on the form of the boundary condition imposed at $r=0$. A natural choice is to impose regularity, i.e.,~$\phi_{\omega\ell} \sim r^{-\lambda+1/2}$ as $r \rightarrow 0$ for $\text{Re} [\lambda] < 0$. An advantage of this choice is that it leads to the same symmetry as in the black hole case for the $S$-matrix elements (see Refs.~\cite{Andersson:1994rk,Folacci:2019cmc}), viz.,
\begin{equation}\label{Matrix_S_CAM_symm}
S_{-\lambda -1/2}(\omega) =  e^{-2i \pi \lambda} \, S_{\lambda -1/2}(\omega)
\end{equation}

\begin{figure}[htb]
\centering
 \includegraphics[scale=0.50]{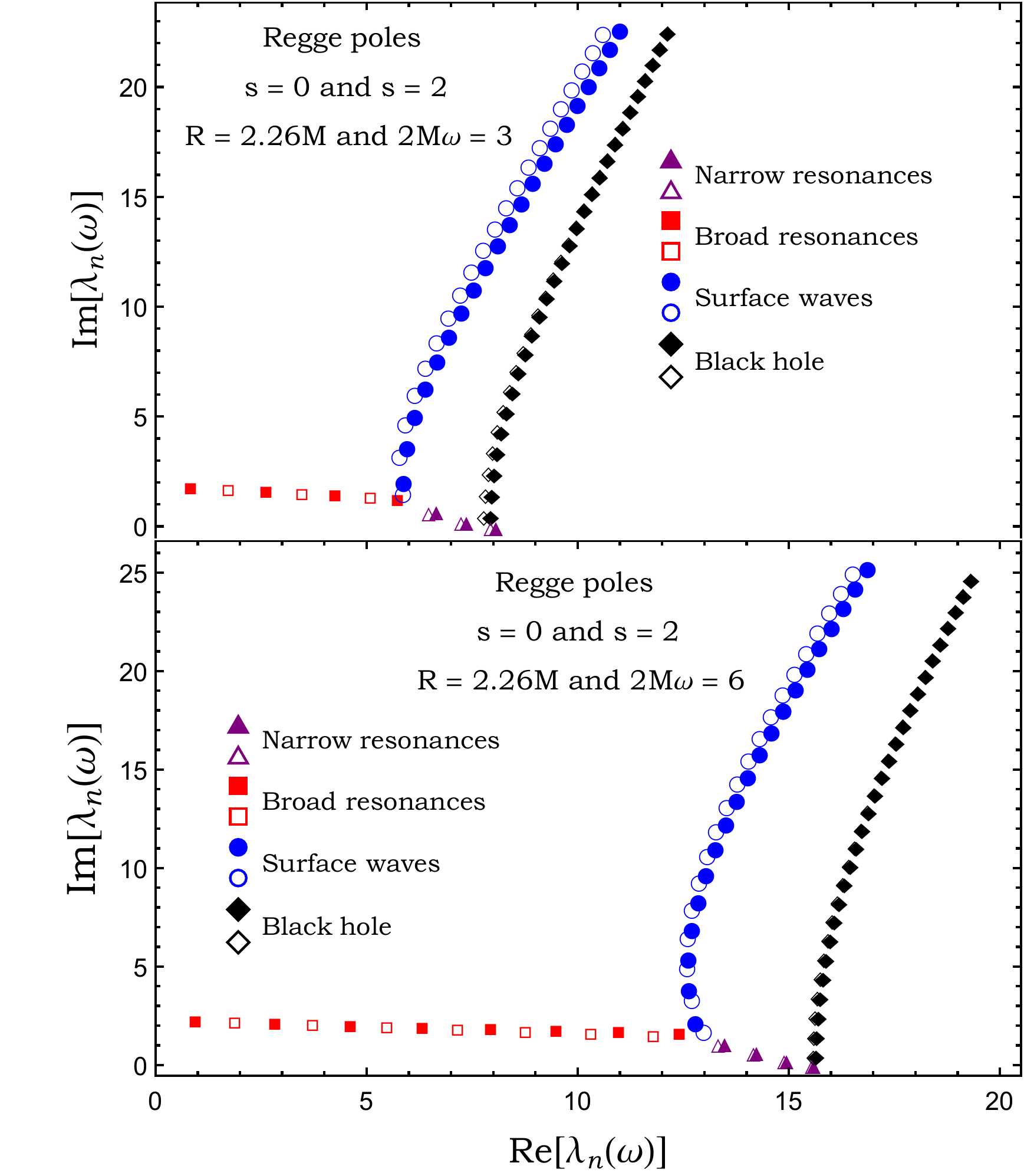}
\caption{\label{RP_R226_approx_2Mw_3_6_s_1}  The Regge poles $\lambda_n(\omega)$ for the scalar field (unfilled markers) and for the axial gravitational waves (filled markers). 
}
\label{fig:rp2}
\end{figure}

Furthermore, since the wave equation is invariant under the transformation $\lambda \rightarrow -\lambda$, it follows that the distribution of poles in the left-half of the complex plane follows by reflection through the origin. Thus, for $\omega > 0$, the 4th and 2nd quadrants are devoid of poles, and the 3rd quadrant has the same structure as shown in Fig.~\ref{fig:rp1} and~\ref{fig:rp2}.
Conversely, for $\omega < 0$, and due to the symmetry relation~\eqref{symmetry_relation}, we may use $\lambda_n(-\omega) = \lambda_n(\omega)^*$ to establish that there are poles in the 2nd and 4th quadrants only.

Figure \ref{fig:rp2} shows the spectrum for a UCO with $R/M = 2.26$. These plots show evidence for an additional branch of modes that emerges from the point where the first two branches meet. The number of modes in this branch increases as the radius of the body approaches the Buchdahl limit $R \rightarrow \tfrac{9}{4}M$.

The Regge pole spectrum for a compact body is qualitatively similar to the Regge pole spectrum found in Mie scattering of electromagnetic waves by a transparent sphere of refractive index $\tilde{n}$. This has been studied since the 1960s; see for example Fig.~9.2 in Ref.~\cite{Nussenzveig:2006}. 
Henceforth we shall adopt the terminology of Nussenzveig \cite{Nussenzveig:2006}, in which the three branches are labeled as:
\begin{enumerate}
 \item \emph{Broad resonances}: approximately uniformly spaced poles above the real axis with approximately constant imaginary part; somewhat sensitive to internal structure ($r < R$);
 \item \emph{Surface waves}: highly damped modes that are relatively insensitive to the internal structure and which depend chiefly on the surface geometry.
 \item \emph{Narrow resonances}: modes approaching the real axis corresponding to trapped modes which only appear in the UCO case ($R < 3M$).
\end{enumerate}
In Figs.~\ref{fig:rp1} and \ref{fig:rp2} the broad resonances, surface waves and narrow resonances are indicated by red squares, blue circles and purple triangles, respectively; and black diamonds indicate the black hole Regge poles. Whereas there exists an infinite number of poles of the surface-wave branch, in principle, there are no narrow-resonance poles at all in the $R/M = 6$ case, and just $\le 4$ narrow-resonance poles in the $R/M = 2.26$ case. The narrow-resonance branch is observed to end close to the start of the black hole branch.

Data for Regge poles $\lambda_n(\omega)$ is listed in Table \ref{tab:table1} (scalar field, $R/M=6$), Table \ref{tab:table2} (scalar field, $R/M = 2.26$), Table \ref{tab:table3} (axial $s=2$, $R/M = 6$) and Table \ref{tab:table4} (axial $s=2$, $R/M = 2.26$). The values are labelled by branch (broad; surface; narrow). For the scalar field, the associated residues (see Eq.~\eqref{residues_RP}) are also presented in Tables \ref{tab:table1} and Table \ref{tab:table2}.

\begingroup
\squeezetable
\begin{table*}[htp]
\begin{threeparttable}[htp]
\caption{\label{tab:table1} The lowest Regge poles $\lambda_{n}(\omega)$ for the scalar field and the associated residues $r_{n}(\omega)$. The radius of the compact bodies is $R = 6M$.}
\smallskip
\centering
\begin{ruledtabular}
\begin{tabular}{cccccc}
 $n$ & $2 M \omega$  & $\lambda^{\text{(S-W)\tnote{1}}}_n(\omega)$ & $\lambda^{\text{(B-R)\tnote{2}}}_n(\omega)$ & $r^{\text{(S-W)}}_{n}(\omega)$ & $r^{\text{(B-R)}}_{n}(\omega)$
 \\ \hline
$1$  & $3$  & $ 9.64850+2.76784 i$  & $1.56219+2.33072 i$  & $-12.41483-0.10424 i$  & $ -0.184457+0.480330 i$    \\
     & $16$  & $56.00945+5.71038 i$  & $ 0.62529+3.27098 i$  & $-447.5395+25.2912 i$  & $-0.322061-0.088002 i $  \\

$2$  & $3$  & $ 10.71986+5.16209 i $  & $3.81484+2.48159 i$  & $13.8486+24.3824 i$  & $0.290952+1.043116 i$    \\
     & $16$  & $ 58.442656+9.18793 i$ & $2.64868+3.31439 i$  & $5188.750-859.909 i$  & $ -0.381581-0.077583 i$    \\

$3$  & $3$  & $11.62296+7.17454 i$  & $ 6.35675+2.64104 i$  & $39.4189-12.3554 i$ & $2.83038-0.28686 i$    \\
     & $16$  & $60.20374+12.14965 i$  & $4.70011+3.35821 i$  & $-29331.71-18578.38 i$  & $-0.456423-0.021249 i$ \\

$4$  & $3$  & $ 12.4297+8.9960 i$  & $/$  & $ 13.2301-50.8802 i$ & $/$    \\
     & $16$  & $ 61.67700+14.84728 i $  & $6.78093+3.40257 i$  & $-15868.9+161199.9 i$  & $-0.528929+0.106794 i$ \\

$5$  & $3$  & $13.1734+10.6929 i$  & $/$  & $-33.7366-51.7404 i$ & $/$     \\
     & $16$  & $62.98626+17.37165 i$  & $8.89270+3.44762 i$  & $589920.5-79507.8 i$  & $-0.550038+0.330275 i$    \\

$6$  & $3$  & $13.8709+12.2989 i$  & $/$   & $-66.4436-20.7767 i$ & $/$    \\
     & $16$  & $ 64.18605+19.76911 i$  & $11.03720+3.49356 i$  & $-360464.-1.797518\times 10^6 i$  & $ -0.426365+0.639191 i$    \\

$7$  & $3$  & $14.5322+13.8342 i$  & $/$   & $-73.0825+21.9088 i$ & $/$     \\
     & $16$  & $65.30640+22.06743 i$  & $13.21653+3.54058 i$  & $-4.880638\times 10^6+646112. i$  & $-0.038292+0.926498 i$    \\

$8$  & $3$  & $15.1640+15.3122 i$  & $/$   & $-56.3641+59.6187 i$ & $/$     \\
     & $16$  & $66.36581+24.28491 i$  & $15.43310+3.5889 i$  & $-479098.+1.1836070\times 10^7 i$  & $0.652285+0.920876 i$    \\

$9$  & $3$  & $15.7709+16.7425 i$  & $/$   & $-25.0183+83.3731 i$ & $/$     \\
     & $16$  & $67.37659+26.43447 i$  & $17.6898+3.6390 i$  & $2.487209\times 10^7+7.72797\times 10^6 i$  & $1.363464+0.248276 i$    \\

$10$  & $3$  & $16.3565+18.1321 i$  & $/$   & $11.7631+90.6815 i$ & $/$     \\
     & $16$  & $68.34738+28.52564 i $  & $19.9900+3.6910 i$  & $3.163822\times 10^7-4.265475\times 10^7 i$  & $1.29469-1.13096 i$    \\
\end{tabular}
\end{ruledtabular}
\begin{tablenotes}
     \item[1] S-W : Surface waves
     \item[2] B-R : Broad resonances
   \end{tablenotes}
\end{threeparttable}
\end{table*}
\endgroup

\begingroup
\squeezetable
\begin{table*}[htp]
\begin{threeparttable}[htp]
\caption{\label{tab:table2} The lowest Regge poles $\lambda_{n}(\omega)$ for the scalar field and the associated residues $r_{n}(\omega)$. The radius of the compact bodies is $R = 2.26M$.}
\smallskip
\centering
\begin{ruledtabular}
\begin{tabular}{cccccccc}
 $n$ & $2M\omega$  & $\lambda^{\text{(S-W)\tnote{1}}}_n(\omega)$  & $\lambda^{\text{(B-R)\tnote{2}}}_n(\omega)$ & $\lambda^{\text{(N-R)\tnote{3}}}_n(\omega)$ & $r^{\text{(S-W)}}_{n}(\omega)$ & $r^{\text{(B-R)}}_{n}(\omega)$ & $r^{\text{(N-R)}}_{n}(\omega)$
 \\ \hline
$1$  & $3$  & $5.871590+1.553799 i$  & $1.73455+1.64951 i  $  & $ 6.48474+0.68765 i $  & $-179.7945+131.4187 i $ & $ -1.52081-2.30968 i$ & $-2.5672-15.3797 i $  \\
     & $6$  & $12.991923+1.754967 i $  & $ 1.89664+2.13696 i  $  & $ 13.34118+1.13496 i $  & $4356.193+647.790 i $ & $  -0.66176-1.31963 i$ & $  -390.218+379.906 i $  \\

$2$  & $3$  & $5.778805+3.228990 i  $  & $ 3.48084+1.45765 i$  & $  7.25606+0.24457 i$  & $428.6893-235.0321 i $ & $16.2123+5.2371 i $ & $ -0.272250-1.150335 i$  \\
     & $6$  & $12.705495+3.383881 i $  & $ 3.74238+2.01309 i $  & $14.18757+0.68182 i $  & $-35075.99-9772.94 i $ & $-2.93679+4.83548 i $ & $ -11.3519+34.5571 i $  \\

$3$  & $3$  & $ 5.924546+4.705899 i $  & $  5.10229+1.29099 i$  & $ 7.95763+0.01764 i$  & $-404.6185-390.8531 i$ & $ 70.4849+54.1888 i  $ & $ -0.0370202-0.0048174 i $  \\
     & $6$  & $12.596259+4.982661 i $  & $  5.49829+1.89576 i $  & $14.9017+0.2912 i $  & $82360.19+81990.53 i$ & $ 6.7872-16.9564 i$ & $0.27028+2.27905 i  $  \\

$4$  & $3$  & $ 6.144986+6.043188 i $  & $ /$  & $/ $  & $ -471.5443+314.3116 i  $ & $ /$ & $/ $  \\
     & $6$  & $ 12.614598+6.503749 i  $  & $ 7.17509+1.78279 i $  & $15.5621+0.0422 i  $  & $39281.5-229393.2 i  $ & $39.6176+33.5152 i  $ & $0.1011154+0.0020569 i $  \\

$5$  & $3$  & $6.398427+7.281723 i $  & $ /$  & $/ $  & $37.8777+546.8945 i $ & $ /$ & $/ $  \\
     & $6$  & $  12.71646+7.95208 i $  & $ 8.78112+1.67243 i  $  & $ /$  & $ -356055.5+34945.9 i$ & $ 2.1175+134.5962 i $ & $/ $  \\

$6$  & $3$  & $ 6.666837+8.447532 i $  & $/ $  & $ /$  & $418.7890+315.4209 i$ & $/ $ & $/ $  \\
     & $6$  & $ 12.87420+9.33552 i $  & $ 10.32300+1.56317 i$  & $/ $  & $45934.6+468157.5 i $ & $ 66.944+324.598 i$ & $/ $  \\

$7$  & $3$  & $ 6.941642+9.557619 i $  & $ /$  & $/ $  & $499.2703-37.6476 i $ & $ /$ & $ /$  \\
     & $6$  & $13.06993+10.66226 i $  & $ 11.80630+1.45720 i  $  & $ /$  & $558619.4+61956.5 i $ & $ 833.855+78.332 i $ & $/ $  \\

$8$  & $3$  & $7.218463+10.623548 i $  & $ /$  & $/ $  & $ 367.2578-307.7533 i  $ & $ /$ & $/ $  \\
     & $6$  & $13.29184+11.93979 i $  & $/ $  & $/ $  & $ 293571.8-559756.8 i $ & $/ $ & $/ $  \\

$9$  & $3$  & $ 7.494953+11.653498 i $  & $/ $  & $/ $  & $147.3038-435.8160 i  $ & $ /$ & $/ $  \\
     & $6$  & $ 13.53197+13.17461 i  $  & $/ $  & $/ $  & $ -376511.5-570254.0 i  $ & $/$ & $ /$  \\

$10$ & $3$  & $7.76982+12.65345 i $  & $/ $  & $/ $  & $-71.8294-437.3469 i  $ & $/ $ & $ /$  \\
     & $6$  & $ 13.78485+14.37216 i$  & $ /$  & $/ $  & $ -719306.1-20011.7 i $ & $ /$ & $/ $  \\

\end{tabular}
\end{ruledtabular}
\begin{tablenotes}
     \item[1] S-W : Surface waves
     \item[2] B-R : Broad resonances
     \item[3] N-R : Narrow  resonances
   \end{tablenotes}
\end{threeparttable}
\end{table*}
\endgroup

\begingroup
\squeezetable
\begin{table}[htp]
\begin{threeparttable}[htp]
\caption{\label{tab:table3} The lowest Regge poles $\lambda_{n}(\omega)$ for the axial gravitational waves. The radius of the compact bodies is $R = 6M$.}
\smallskip
\centering
\begin{ruledtabular}
\begin{tabular}{cccc}
 $n$ & $2M\omega$  & $\lambda^{\text{(S-W)\tnote{1}}}_n(\omega)$ & $\lambda^{\text{(B-R)\tnote{2}}}_n(\omega)$
 \\ \hline
$1$  & $3$  & $10.004639+2.935907 i$  & $0.461101+2.2269826 i $   \\
     & $16$ & $56.179459+5.874240 i $ & $1.624880+3.2909174 i $
 \\

$2$  & $3$  & $11.205047+5.343083 i$  & $2.551455+2.401243 i $     \\
     & $16$ & $58.717121+9.365689 i  $& $3.659398+3.336427 i $
 \\

$3$  & $3$  & $12.166772+7.344219 i $  & $4.790413+2.635624 i  $     \\
     & $16$ & $60.528800+12.332399 i $ & $5.722341+3.384112 i $
 \\

$4$  & $3$  & $13.009174+9.155489 i  $  & $7.164229+2.930135 i $     \\
     & $16$ & $62.040466+15.032415 i  $ & $7.815081+3.434319 i  $
      \\

$5$  & $3$  & $13.777754+10.843971 i $  & $/ $   \\
     & $16$ & $63.379530+17.557778 i $  & $9.939189+3.487474 i $      \\

$6$  & $3$  & $14.494020+12.442853 i $  & $/ $      \\
     & $16$ & $64.603684+19.955393 i $  & $12.096463+3.544109 i  $
   \\

$7$  & $3$  & $15.170211+13.972029 i $  & $ /$       \\
     & $16$ & $65.744593+22.253317 i $  & $14.288975+3.604892 i $
         \\

$8$  & $3$  & $15.814162+15.444717 i $  & $/ $        \\
     & $16$ & $66.821746+24.470040 i $  & $16.519109+3.670678 i $
         \\

$9$  & $3$  & $16.431296+16.870298 i $  & $/ $      \\
     & $16$ & $67.848089+26.618578 i $  & $18.789625+3.742581 i $
         \\

$10$  & $3$  & $17.025584+18.255740 i $  & $/ $     \\
     & $16$  & $68.832711+28.708547 i  $  & $21.103712+3.822080 i $
        \\
\end{tabular}
\end{ruledtabular}
\begin{tablenotes}
     \item[1] S-W : Surface waves
     \item[2] B-R : Broad resonances
   \end{tablenotes}
\end{threeparttable}
\end{table}
\endgroup

\begingroup
\squeezetable
\begin{table}[htp]
\fontsize{5.}{10}
\begin{threeparttable}[htp]
\captionsetup{font=small}
\caption{\label{tab:table4} The lowest Regge poles $\lambda_{n}(\omega)$ for the axial gravitational waves. The radius of the compact bodies is $R = 2.26M$.}
\smallskip
\centering
\begin{ruledtabular}
\begin{tabular}{ccccc}
 $n$ & $2M\omega$  & $\lambda^{\text{(S-W)\tnote{1}}}_n(\omega)$  & $\lambda^{\text{(B-R)\tnote{2}}}_n(\omega)$ & $\lambda^{\text{(N-R)\tnote{3}}}_n(\omega)$
 \\ \hline
$1$  & $3$  & $5.884755+2.047850 i $  & $0.840822+1.728755 i   $  & $8.0740924+0.022432 i   $   \\
     & $6$  & $12.796673+2.200987 i  $  & $0.959779+2.193488 i   $  & $14.247709+0.699340 i  $    \\

$2$  & $3$  & $5.960332+3.632211 i   $  & $2.633309+1.559084 i   $  & $7.386249+0.264972 i   $  \\
     & $6$  & $12.640961+3.871470 i  $  & $2.842316+2.075844 i  $  & $14.969378+0.297175 i  $  \\

$3$  & $3$  & $6.153107+5.055691 i   $  & $4.262509+1.419601 i   $  & $6.662827+0.729518 i  $   \\
     & $6$  & $ 12.628734+5.440977 i$  & $4.629152+1.973668 i  $  & $13.496329+1.146922 i  $  \\

$4$  & $3$  & $6.403231+6.358703 i  $  & $ 5.746584+1.200761 i$  & $/ $   \\
     & $6$  & $12.711478+6.932783 i  $  & $6.330877+1.883529 i  $  & $15.612017+0.049364 i  $   \\

$5$  & $3$  & $6.678538+7.572564 i $  & $ /$  & $/ $   \\
     & $6$  & $12.859039+8.354930 i   $  & $7.954922+1.803188 i    $  & $ /$  \\

$6$  & $3$  & $6.964301+8.719285 i  $  & $/ $  & $ /$  \\
     & $6$  & $13.050962+9.715667 i  $  & $9.505808+1.730884 i  $  & $/ $   \\

$7$  & $3$  & $7.253473+9.813908 i   $  & $ /$  & $/ $  \\
     & $6$  & $13.273382+11.022841 i  $  & $10.985288+1.664216 i   $  & $ /$   \\

$8$  & $3$  & $7.542524+10.866909 i $  & $ /$  & $/ $   \\
     & $6$  & $13.516855+12.283449 i $  & $12.423672+1.572865 i $  & $/ $   \\

$9$  & $3$  & $7.829631+11.885800 i   $  & $/ $  & $/ $   \\
     & $6$  & $13.774891+13.503511 i  $  & $/ $  & $/ $   \\

$10$ & $3$  & $8.113848+12.876130 i $  & $/ $  & $/ $    \\
     & $6$  & $14.042970+14.688113 i $  & $ /$  & $/ $   \\
\end{tabular}
\end{ruledtabular}
\begin{tablenotes}
     \item[1] S-W : Surface waves
     \item[2] B-R : Broad resonances
     \item[3] N-R : Narrow  resonances
   \end{tablenotes}
\end{threeparttable}
\end{table}
\endgroup

 \subsection{The WKB approximation}\label{subsec:WKB}

To investigate the relationship between the qualitative features of the effective potential (Fig.~\ref{fig:Veff}) and the three branches of Regge poles revealed in Sec.~\ref{subsec:results1}, we now employ the WKB method, with a view to obtaining an approximation that is valid at high frequencies ($M\omega \rightarrow \infty$).

Here we follow the approach of Zhang, Wu and Leung~\cite{Zhang:2011pq}, who applied the WKB method to determine the axial \textit{w}-modes of a variety of stellar models (see also \cite{Volkel:2019gpq}). We adapt their method to obtain analytical approximations for the ``broad resonances'' for massless waves on a stellar background. The starting point is the radial equation (\ref{H_Radial_equation}) with either the effective potential for the scalar field (\ref{Inside_Potentiel}), or for axial gravitational perturbations (\ref{RW_Potentiel_axialGW}).
It is valid only for models with $R/M>3$.

Regge poles and quasinormal modes for relativistic stellar models (of which \textit{w}-modes are a subcategory for gravitational perturbations) both satisfy the same wave equation and the same boundary conditions, but with different interpretations for the angular momentum index and the frequency. Both types of pole satisfy the regularity condition at the origin~\eqref{bc_1_in} and the condition of a purely outgoing wave in the far field
\begin{equation}\label{bc_pole_inf}
\phiout (r) \scriptstyle{\underset{r_\ast \to +\infty}{\sim}}
\displaystyle{  A^{(+)}_{\lambda-1/2} (\omega) e^{+i\omega r_\ast}}.
\end{equation}
Thus, the Regge poles are solutions of Eq.~\eqref{H_Radial_equation} for which the  Wronskian of $\phireg$ and $\phiout$ vanishes (\textit{i.e.}, $A^{(-)}_{\lambda_n(\omega)-1/2} (\omega)=0$), viz.,
\begin{equation}\label{Wronskian}
  W[\phireg, \phiout ]= 0 .
\end{equation}
It has been shown \cite{Zhang:2011pq} that asymptotic expressions of $\phireg$ and $\phiout$ can be derived in asymptotic regions, by using a WKB approximation.

In the interior of the star, the radial function $f(r)$ is
\begin{equation}\label{eq:f_exp}
f(r) = f_0 \left(1 + O\left(\frac{M r^2}{R^3} \right) \right),
\end{equation}
where $f_0$ is a constant, and $h(r) = 1 - 2Mr^2/R^3$. In the vicinity of $r=0$, we may solve Eq.~(\ref{eq:tortoise}) to obtain the tortoise coordinate $r_\ast(r)$, or inversely,
\begin{equation}
r = \sqrt{f_0} \, r_\ast \left(1 + O\left(\frac{M r_\ast^2}{R^3} \right) \right) . \label{eq:rrstar}
\end{equation}
Now let us consider the radial equation (\ref{H_Radial_equation}) in the high-frequency regime with $\ell \gg 1$, such that we may neglect all but the angular momentum terms in the effective potential (\ref{Inside_Potentiel}). Inserting (\ref{eq:rrstar}) and neglecting the quadratic corrections leads to a comparison equation
\begin{equation}
\left[\frac{d^{2}}{dr_{\ast}^{2}}+\omega^{2} - \frac{\ell (\ell+1)}{r_\ast^2} \right] \phi_{\omega\ell}= 0,
\end{equation}
with the regular interior solution
\begin{equation}\label{Approx_origin}
  \phireg \underset{\omega \to \infty}{=} \omega r_* j_{\lambda-1/2} (\omega r_*),
\end{equation}
where $j_{\lambda-1/2}(\cdot)$ is the spherical Bessel function of the first kind.

Near the surface of the body, and in the exterior region, we use the asymptotic forms \cite{Zhang:2011pq}
\begin{equation}
\label{Approx_infinity}
\phiout  \scriptstyle{\underset{\omega \to \infty}{=}}
\left\{
\begin{aligned}
&e^{i\omega(r_*-R_*)}+\mathcal{R} e^{-i\omega (r_*-R_*)}& \scriptstyle{1/\omega\leq r_{*} < R_*,}\\
&\left(1+\mathcal{R}\right) e^{i\omega(r_*-R_*)}        & \scriptstyle{R_*\leq r_{*}<\infty.}
\end{aligned}
\right.
\end{equation}
Here $R_*$ is the tortoise coordinate at the surface of the body,
and $\mathcal{R}$ is a reflection coefficient with the definition given in Ref.~\cite{Berry_1982}.
Because the potential has a direct discontinuity at the surface of the compact body (see. Refs \cite{Zhang:2011pq,Berry_1982} for more details), we have for our model
\begin{equation}\label{R_reflection_1}
  \mathcal{R} = \alpha \omega^{-2}
\end{equation}
with
\begin{eqnarray} \label{eq:alpha}
  \alpha &=& \frac{1}{4} \Delta V(R) = \pm \frac{3M(R-2M)}{4 R^4} ,
\end{eqnarray}
where $\Delta V$ is the discontinuity in the effective potential at the surface, defined in Eq.~(\ref{def:DeltaV}), and the $+$ ($-$) sign denotes the scalar (axial gravitational wave) case.

Inserting the high-frequency approximation for $j_{\lambda-1/2} (\omega r_*)$ into Eq.~\eqref{Approx_origin} we obtain \cite{nist},
\begin{equation}\label{Approx_origin_bis}
  \phi^{}_{\omega,\lambda-1/2} \underset{\omega \to \infty}{\approx} - \sin\left(\frac{\left(\lambda-1/2\right)\pi}{2}-\omega r_*\right).
\end{equation}
Substituting Eqs.~\eqref{Approx_infinity} and \eqref{Approx_origin_bis} into condition \eqref{Wronskian} leads to
\begin{equation}\label{WKB_1}
   e^{ i\pi (\lambda-1/2)- 2 i \omega R_*}=-\mathcal{R}.
\end{equation}
We then solve Eq.~\eqref{WKB_1} to obtain the approximate Regge pole solution
\begin{equation}\label{lambda_Approx}
  \lambda_n \approx \frac{2 \omega R_*}{\pi}-\left(2n \pm \frac{1}{2}\right)+\frac{2 i}{\pi} \ln\left(\frac{2 R^2 \omega}{\sqrt{3M (R-2M)}}\right) .
\end{equation}
This corresponds to a series of Regge poles with spacing $|\Delta \lambda_n|\approx 2$ and an almost-constant imaginary part; these are the broad resonances shown in Figs.~\ref{fig:rp1} and \ref{fig:rp2}.
The formula also correctly accounts for the alternating sequence of the scalar-field and axial-GW modes.

The overtones are labeled by $n =1,2,\ldots$ and the condition $\text{Re} \, \lambda_n > 0$ leads to an upper limit for $n$ of %
\begin{equation}\label{Limit_Broad_RP}
n \leq \left\lfloor \frac{\omega R_*}{\pi} \mp \frac{1}{4}\right\rfloor.
\end{equation}
In other words, there are a finite number of the broad resonances in the first quadrant.

As $M/R \rightarrow 0$ (a large dilute star), the poles move closer to the real axis. When $M/R=0$ (i.e. no star), the potential is $C^{\infty}$ and the branch does not exist.

For any stellar model (such as a polytropic fluid sphere), the potential may in general be $C^{N}$ at the surface (and $C^{\infty}$ elsewhere). In this case, $\mathcal{R}$ and thus the broad resonance branch depends on the discontinuity in the $N$-th derivative of the potential at $r=R$. It can be shown that the imaginary part of the poles is proportional to $(N+1)$ (see Eq.~(4.5) of Ref. \cite{Zhang:2011pq}), so in the large-$N$ limit the associated modes have large imaginary parts and little physical consequence.

Recall that the above analysis is not appropriate for compact bodies with $R/M<3$ and black holes (since the effective potential has a qualitatively different structure, and the ansatz Eq.~(\ref{Approx_infinity}) is not valid).

Table \ref{tab:table5} compares the numerically-determined Regge poles with the WKB approximation in Eq.~(\ref{lambda_Approx}),  for the scalar-field case. The data shows that, while the leading-order WKB approximation captures the essential features of the broad resonance branch, it is not particularly accurate. Higher-order extensions are possible in principle, but not pursued here.


\begingroup
\squeezetable
\begin{table}[htp]
\caption{\label{tab:table5} The lowest Regge poles $\lambda_{n}(\omega)$ for the scalar field versus WKB results given by Eq.~\eqref{lambda_Approx}. The radius of the compact bodies is $R = 6M$.}
\smallskip
\centering
\begin{ruledtabular}
\begin{tabular}{cccc}
 $n$ & $2 M \omega$  & $\lambda^{\text{(B-R)}}_n(\omega)$ & $\lambda^{\text{(B-R, WKB)}}_n(\omega)$
 \\ \hline
$1$  & $3$  & $1.56219+2.33072 i$  & $1.592793+2.189767i$   \\
     & $16$ & $ 0.62529+3.27098 i$ & $0.661564+3.255453i $   \\

$2$  & $3$  & $3.81484+2.48159 i$  & $3.592793+2.189767i $    \\
     & $16$ & $2.64868+3.31439 i$  & $2.661564+3.255453i $    \\

$3$  & $3$  & $6.35675+2.64104 i$ & $5.592793+2.189767i $    \\
     & $16$ & $4.70011+3.35821 i$  & $4.661564+3.255453i $    \\

$4$  & $3$  & $/$  & $ / $   \\
     & $16$ & $6.78093+3.40257 i$  & $6.661564+3.255453i $  \\

$5$  & $3$  & $/$  & $/$     \\
     & $16$ & $8.89270+3.44762 i$  & $8.661564+3.255453i $   \\

$6$  & $3$  & $/$  & $/$    \\
     & $16$ & $11.03720+3.49356 i$ & $10.661564+3.255453i $ \\

$7$  & $3$  & $/$ & $/$     \\
     & $16$ & $13.21653+3.54058 i$  & $12.661564+3.255453i$ \\

$8$  & $3$ & $/$   & $/$     \\
     & $16$ & $15.4331+3.5889 i$  & $14.661564+3.255453i $  \\

$9$  & $3$ & $/$   & $/$   \\
     & $16$ & $17.6898+3.6390 i$  & $16.661564+3.384517i $   \\

$10$  & $3$  & $/$  & $/$   \\
     & $16$  & $19.9900+3.6910 i$  & $18.661564+3.255453i $  \\
\end{tabular}
\end{ruledtabular}
\end{table}
\endgroup

\section{Scattering and CAM theory} \label{sec:CAM}
In this section we calculate the scattering cross section $d\sigma / d\Omega$ by means of the CAM method, and we compare with results obtained in the standard way from a partial-wave series \cite{Dolan:2017rtj}.

\subsection{The partial wave expansion} \label{subsec:partial}

For a scalar field, the differential scattering cross section is given by
\begin{equation}\label{Scalar_Scattering_diff}
  \frac{d\sigma}{d\Omega} = |\hat{f}(\omega,\theta)|^2
\end{equation}
where
\begin{equation}\label{Scalar_Scattering_amp}
 \hat{f}(\omega,\theta) = \frac{1}{2 i \omega} \sum_{\ell = 0}^{\infty} (2\ell+1)[S_{\ell}(\omega)-1]P_{\ell}(\cos\theta)
\end{equation}
is the scattering amplitude (for details see e.g.~\cite{Dolan:2017rtj} and references therein).  In Eq.~(\ref{Scalar_Scattering_amp}), the functions $P_{\ell}(\cos\theta)$ are the Legendre polynomials \cite{AS65}, and the $S$-matrix elements $S_{\ell}(\omega)$ appearing in Eq.~(\ref{Scalar_Scattering_amp}) were defined in Eq.~(\ref{Matrix_S}).

\subsection{CAM representation of the scattering amplitude}
\label{SecIIc}


To construct the CAM representation of $\hat{f}(\theta)$, we follow the steps in section~II of Ref.~\cite{Folacci:2019cmc} and recall the main results below.

The Sommerfeld-Watson transformation \cite{Watson18,Sommerfeld49,Newton:1982qc} permits us to replace a sum with an integral, viz.,
\begin{equation}\label{SWT_gen}
\sum_{\ell=0}^{+\infty} (-1)^\ell F(\ell)= \frac{i}{2} \int_{\cal C} d\lambda \, \frac{F(\lambda -1/2)}{\cos (\pi \lambda)} ,
\end{equation}
where $F(\cdot)$ is any function without singularities on the real $\lambda$ axis. Applying this to Eq.~(\ref{Scalar_Scattering_amp}) allows us to replace the discrete sum over the ordinary angular momentum $\ell$ with a contour integral in the complex $\lambda$ plane (that is, in the complex $\ell$ plane with $\lambda = \ell +1/2$). By noting that $P_\ell (\cos \theta)=(-1)^\ell P_\ell (-\cos \theta)$, we obtain
\begin{eqnarray}\label{SW_Scalar_Scattering_amp}
& & \hat{f}(\omega,\theta) = \frac{1}{2 \omega}  \int_{\cal C} d\lambda \, \frac{\lambda}{\cos (\pi \lambda)} \nonumber \\
&&  \qquad\qquad   \times \left[ S_{\lambda -1/2} (\omega) -1 \right]P_{\lambda -1/2} (-\cos \theta).
\end{eqnarray}
In Eqs.~(\ref{SWT_gen}) and (\ref{SW_Scalar_Scattering_amp}), the integration contour encircles counterclockwise the positive real axis of the complex $\lambda$ plane, i.e., we take ${\cal C}=]+\infty +i\epsilon,+i\epsilon] \cup
[+i\epsilon,-i\epsilon] \cup [-i\epsilon, +\infty -i\epsilon[$ with $\epsilon \to 0_+$ (see Fig.~1 in Ref~\cite{Folacci:2019cmc}).

The Legendre function of the first kind $P_{\lambda -1/2} (z)$ denotes the analytic extension of the Legendre polynomials $P_\ell (z)$. It is defined in terms of hypergeometric functions by \cite{AS65}
\begin{equation}\label{Def_ext_LegendreP}
P_{\lambda -1/2} (z) = F[1/2-\lambda,1/2+\lambda;1;(1-z)/2].
\end{equation}
In Eq.~(\ref{SW_Scalar_Scattering_amp}), $S_{\lambda -1/2} (\omega)$ denotes ``the'' analytic extension of $S_\ell (\omega)$. It is given by [see Eq.~(\ref{Matrix_S})]
\begin{equation}\label{Matrix_S_CAM}
  S_{\lambda -1/2}(\omega) =  e^{i(\lambda + 1/2)\pi} \, \frac{A_{\lambda -1/2}^{(+)}(\omega)}{A_{\lambda -1/2}^{(-)}(\omega)}
\end{equation}
where the complex amplitudes $A^{(-)}_{\lambda -1/2} (\omega)$ and  $A^{(+)}_{\lambda -1/2} (\omega)$ are defined from the analytic extension of the modes $\phi_{\omega \ell}$, i.e., from the function $\phi_{\omega ,\lambda -1/2}$.

It is also important to recall that the poles of  $S_{\lambda-1/2}(\omega)$ in the complex $\lambda$ plan (i.e., the Regge poles) and which are lying in the first and third quadrants, symmetrically distributed with respect to the origin $O$, are defined  as  the zeros of the coefficient  $A^{(-)}_{\lambda-1/2} (\omega)$ [see Eq.~(\ref{Matrix_S_CAM})], i.e., the values $\lambda_n(\omega)$ such that
\begin{equation}\label{PR_def_Am}
A^{(-)}_{\lambda_n(\omega)-1/2} (\omega)=0,
\end{equation}
with $n=1,2,3,\ldots$.

The residue of the matrix $S_{\lambda-1/2}(\omega)$ at the pole $\lambda=\lambda_n(\omega)$ is defined by [see Eq.~(\ref{Matrix_S_CAM})]
\begin{equation}\label{residues_RP}
r_n(\omega)=e^{i\pi [\lambda_n(\omega)+1/2]} \left[ \frac{A_{\lambda -1/2}^{(+)}(\omega)}{\frac{d}{d \lambda}A_{\lambda -1/2}^{(-)}(\omega)}\right]_{\lambda=\lambda_n(\omega)}.
\end{equation}
These residues play a central role in the complex angular momentum paradigm.

Now, we ``deform'' the contour ${\cal C}$ in Eq.~(\ref{SW_Scalar_Scattering_amp}) in order to collect, by using Cauchy's theorem, the Regge poles contributions. This is achieved by following, \textit{mutatis mutandis}, the approach developed in Ref~\cite{Folacci:2019cmc} (see more particularly Sec. IIB~3 and Fig.~1). We obtain
\begin{equation}\label{CAM_Scalar_Scattering_amp_tot}
\hat{f} (\omega, \theta) =  \hat{f}^\text{\tiny{B}} (\omega, \theta) +  \hat{f}^\text{\tiny{RP}} (\omega, \theta)
\end{equation}
where
\begin{subequations}\label{CAM_Scalar_Scattering_amp_decomp}
\begin{equation}\label{CAM_Scalar_Scattering_amp_decomp_Background}
\hat{f}^\text{\tiny{B}} (\omega, \theta) = \hat{f}^\text{\tiny{B},\tiny{Re}} (\omega, \theta)+ \hat{f}^\text{\tiny{B},\tiny{Im}} (\omega, \theta)
\end{equation}
is a background integral contribution with
\begin{equation}\label{CAM_Scalar_Scattering_amp_decomp_Background_a}
\hat{f}^\text{\tiny{B},\tiny{Re}} (\omega, \theta) = \frac{1}{\pi \omega} \int_{{\cal C}_{-}} d\lambda \, \lambda S_{\lambda -1/2}(\omega) Q_{\lambda -1/2}(\cos \theta +i0)
\end{equation}
and
\begin{eqnarray}\label{CAM_Scalar_Scattering_amp_decomp_Background_b}
\hat{f} && ^\text{\tiny{B},\tiny{Im}}  (\omega, \theta) = \frac{1}{2 \omega}\left(\int_{+i\infty}^{0} d\lambda \, \left[S_{\lambda -1/2}(\omega) P_{\lambda-1/2} (-\cos \theta) \right. \right.\nonumber \\
 - && \left.  \left. S_{-\lambda -1/2}(\omega) e^{i \pi \left(\lambda+1/2\right)}P_{\lambda-1/2} (\cos \theta) \right] \frac{\lambda}{\cos (\pi \lambda) } \right).
\end{eqnarray}
\end{subequations}
The second term in Eq.~(\ref{CAM_Scalar_Scattering_amp_tot}) ,
\begin{eqnarray}\label{CAM_Scalar_Scattering_amp_decomp_RP}
& & \hat{f}^\text{\tiny{RP}} (\omega, \theta) = -\frac{i \pi}{\omega}    \sum_{n=1}^{+\infty}   \frac{ \lambda_n(\omega) r_n(\omega)}{\cos[\pi \lambda_n(\omega)]}  \nonumber \\
&&  \qquad\qquad \qquad\qquad \times  P_{\lambda_n(\omega) -1/2} (-\cos \theta),
\end{eqnarray}
is a sum over the Regge poles lying in the first quadrant of the CAM plane. Of course, Eqs.~(\ref{CAM_Scalar_Scattering_amp_tot}), (\ref{CAM_Scalar_Scattering_amp_decomp}) and (\ref{CAM_Scalar_Scattering_amp_decomp_RP}) provide an exact representation of the scattering amplitude $\hat{f} (\omega, \theta)$ for the scalar field, equivalent to the initial partial wave expansion (\ref{Scalar_Scattering_amp}). From this CAM representation, we can extract the contribution $\hat{f}^\text{\tiny{RP}} (\omega, \theta)$ given by (\ref{CAM_Scalar_Scattering_amp_decomp_RP}) which, as a sum over Regge poles, is only an approximation of $\hat{f} (\omega, \theta)$, and which provides us with a corresponding approximation of the differential scattering cross section via (\ref{Scalar_Scattering_diff}).

\subsection{Computational methods}
\label{SecIIIa}

To construct the scattering amplitude \eqref{Scalar_Scattering_amp}, the background integrals~\eqref{CAM_Scalar_Scattering_amp_decomp_Background_a} and~\eqref{CAM_Scalar_Scattering_amp_decomp_Background_b}, and the Regge pole amplitude~\eqref{CAM_Scalar_Scattering_amp_decomp_RP}, we use, \textit{mutatis mutandis}, the computational methods of Refs~\cite{Folacci:2019cmc,Folacci:2019vtt}. In these works the authors calculated the CAM representation of scattering amplitudes for scalar, electromagnetic and gravitational waves by Schwarzschild black hole (see also Ref~\cite{Dolan:2017rtj}).

Due to the long rang nature of the field propagating on the Schwarzschild black hole (outside the compact body), the scattering amplitude~\eqref{Scalar_Scattering_amp} and the background integral~\eqref{CAM_Scalar_Scattering_amp_decomp_Background_a} both suffer a lack of convergence. To overcome this problem, \textit{i.e.}, to accelerate the convergence of this sum and integral, we have used the method described in the Appendix of Ref~\cite{Folacci:2019cmc}. We have performed all the numerical calculations by using {\it Mathematica}. 

\subsection{Numerical results: Scattering cross sections}
\label{subsec:results2}

\begin{figure*}[htp!]
 \includegraphics[scale=0.50]{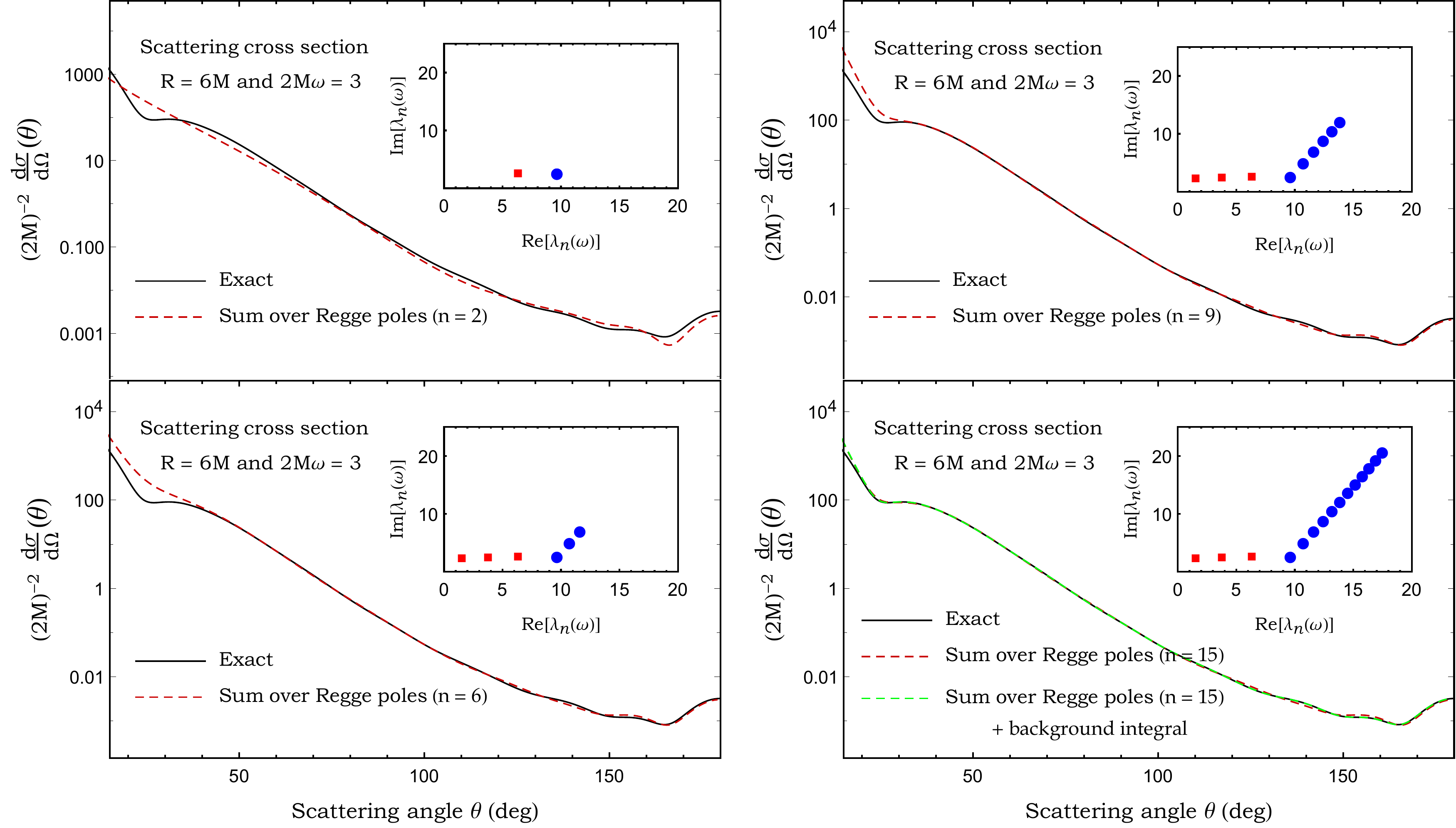}
  \vspace*{-0.35cm}
\caption{\label{S_0_R_6_2Mw_3_Exact_vs_CAM} The scalar cross section of a compact bodies for $2M\omega=3$ and $R=6M$, its Regge pole approximation and the background integral contribution. The plots show the effect of including successively more Regge poles (plots 1--3). In the final plot, the background integral is added, giving a cross section which agrees well with the (regularized) partial-wave sum.}
\end{figure*}

\begin{figure*}[htp!]
 \includegraphics[scale=0.50]{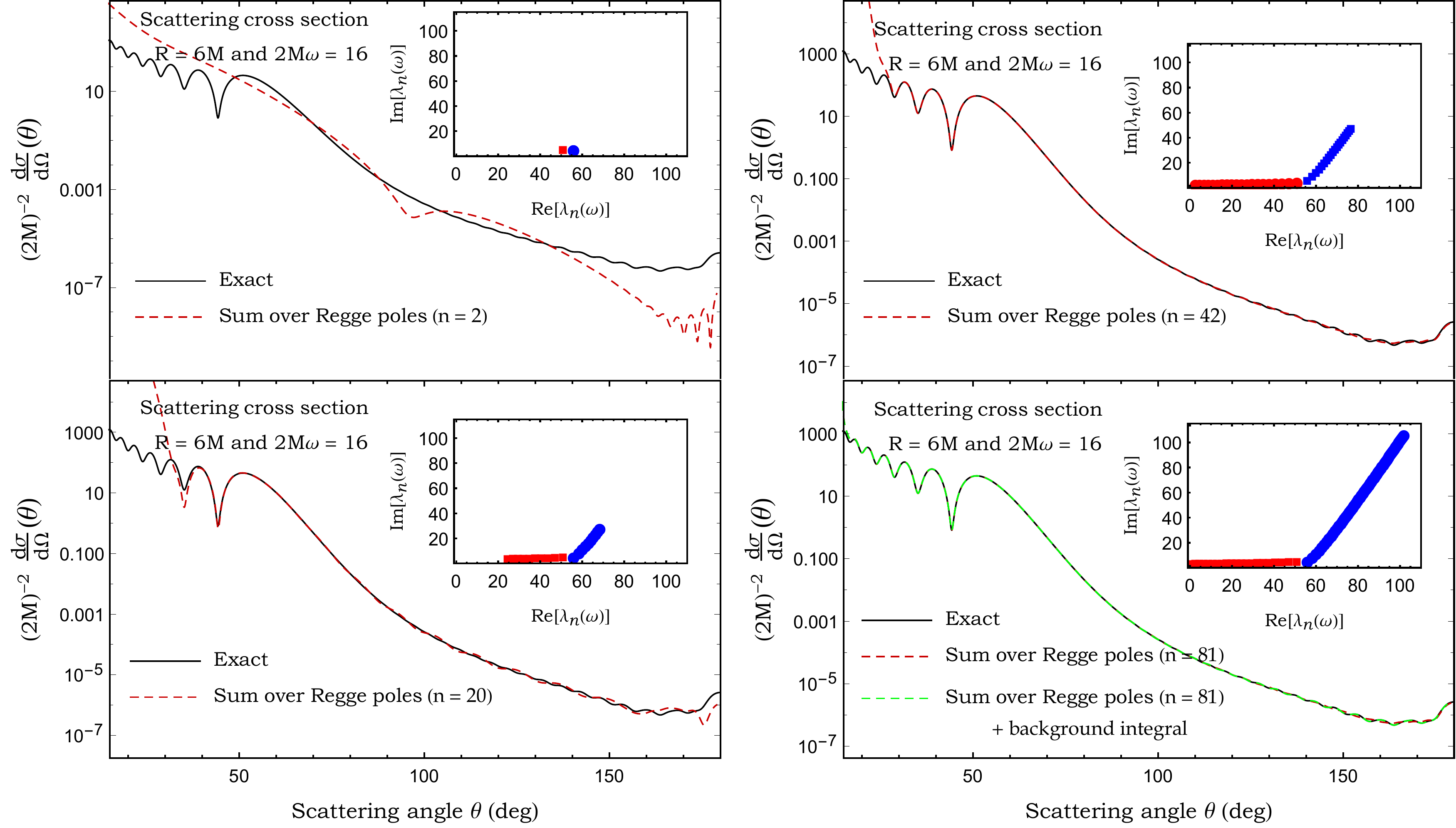}
 \vspace*{-0.35cm}
\caption{\label{S_0_R_6_2Mw_16_Exact_vs_CAM} The scalar cross section of a compact bodies for $2M\omega=16$ and $R=6M$, its Regge pole approximation and the background integral contribution. The plots show the effect of including successively more Regge poles (plots 1--3). In the final plot, the background integral is added, giving a cross section which agrees well with the (regularized) partial-wave sum.}
\end{figure*}

\begin{figure*}[htp!]
\centering
 \includegraphics[scale=0.50]{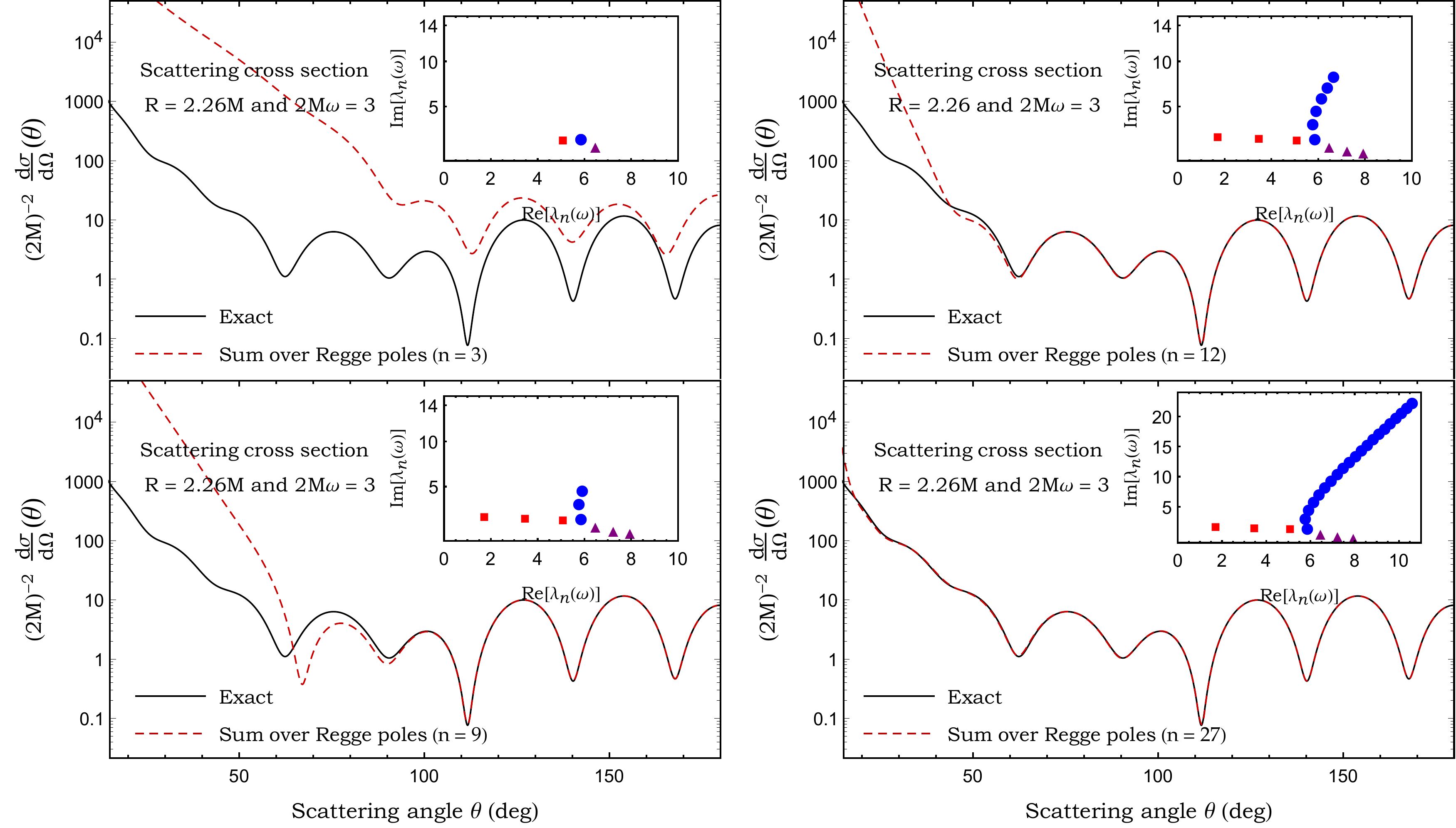}
  \vspace*{-0.45cm}
\caption{\label{S_0_R_2-dot-26_2Mw_3_Exact_vs_CAM} The scalar cross section of an UCOs for $2M\omega=3$ and $R=2.26M$ and its Regge pole approximation. The plots show the effect of including successively more Regge poles (plots 1--3). In the final plot, the sum over Regge poles gives a cross section which agrees well with the (regularized) partial-wave sum for intermediate and large values of the scattering angle.}
\end{figure*}

\begin{figure*}[htp!]
\centering
 \includegraphics[scale=0.50]{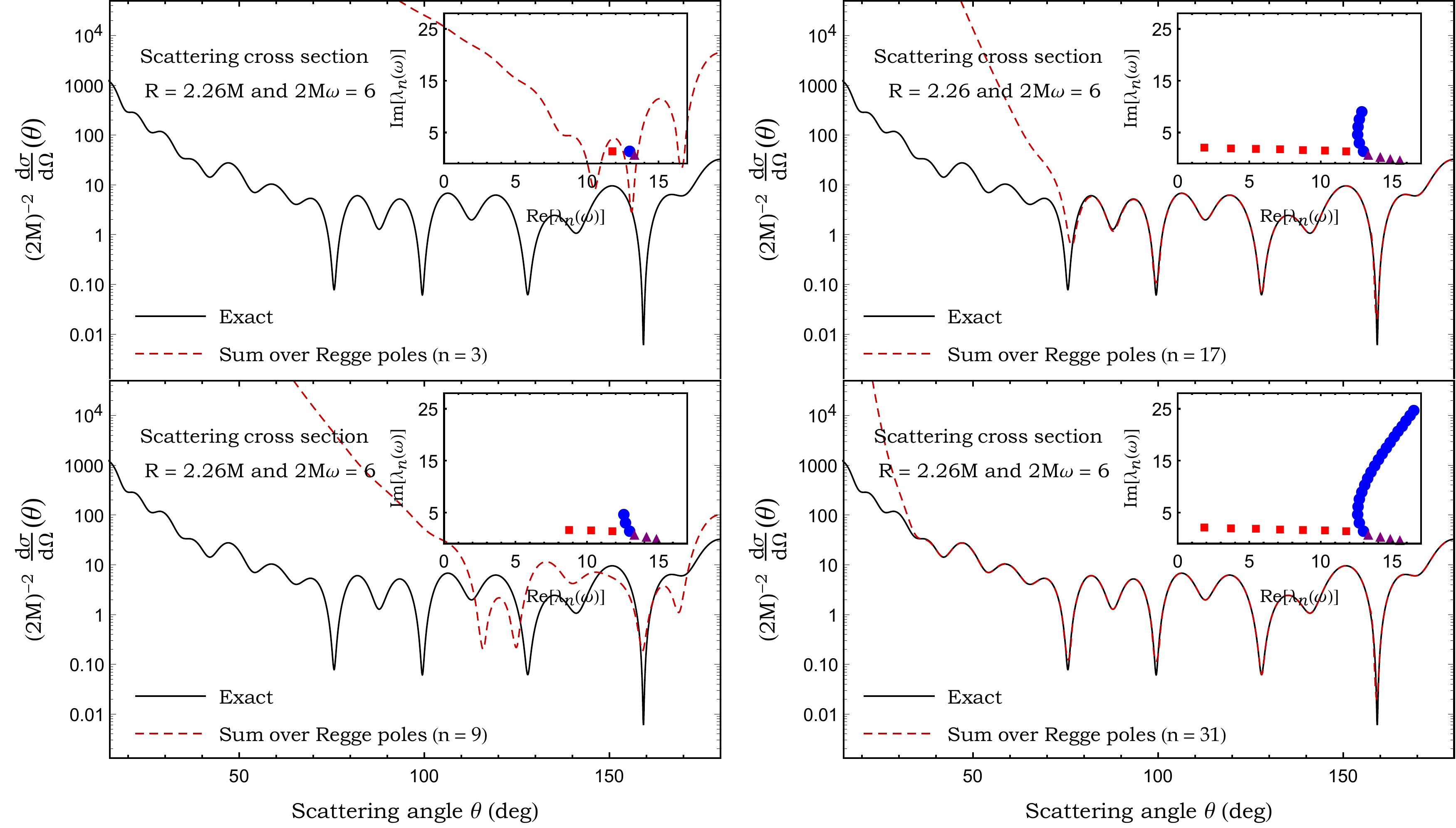}
  \vspace*{-0.45cm}
\caption{\label{S_0_R_2-dot-26_2Mw_6_Exact_vs_CAM} The scalar cross section of an UCOs for $2M\omega=6$ and $R=2.26M$ and its Regge pole approximation. The plots show the effect of including successively more Regge poles (plots 1--3). In the final plot, the sum over Regge poles gives a cross section which agrees well with the (regularized) partial-wave sum for intermediate and large values of the scattering angle.}
\end{figure*}

\begin{figure}[htp!]
\centering
 \includegraphics[scale=0.50]{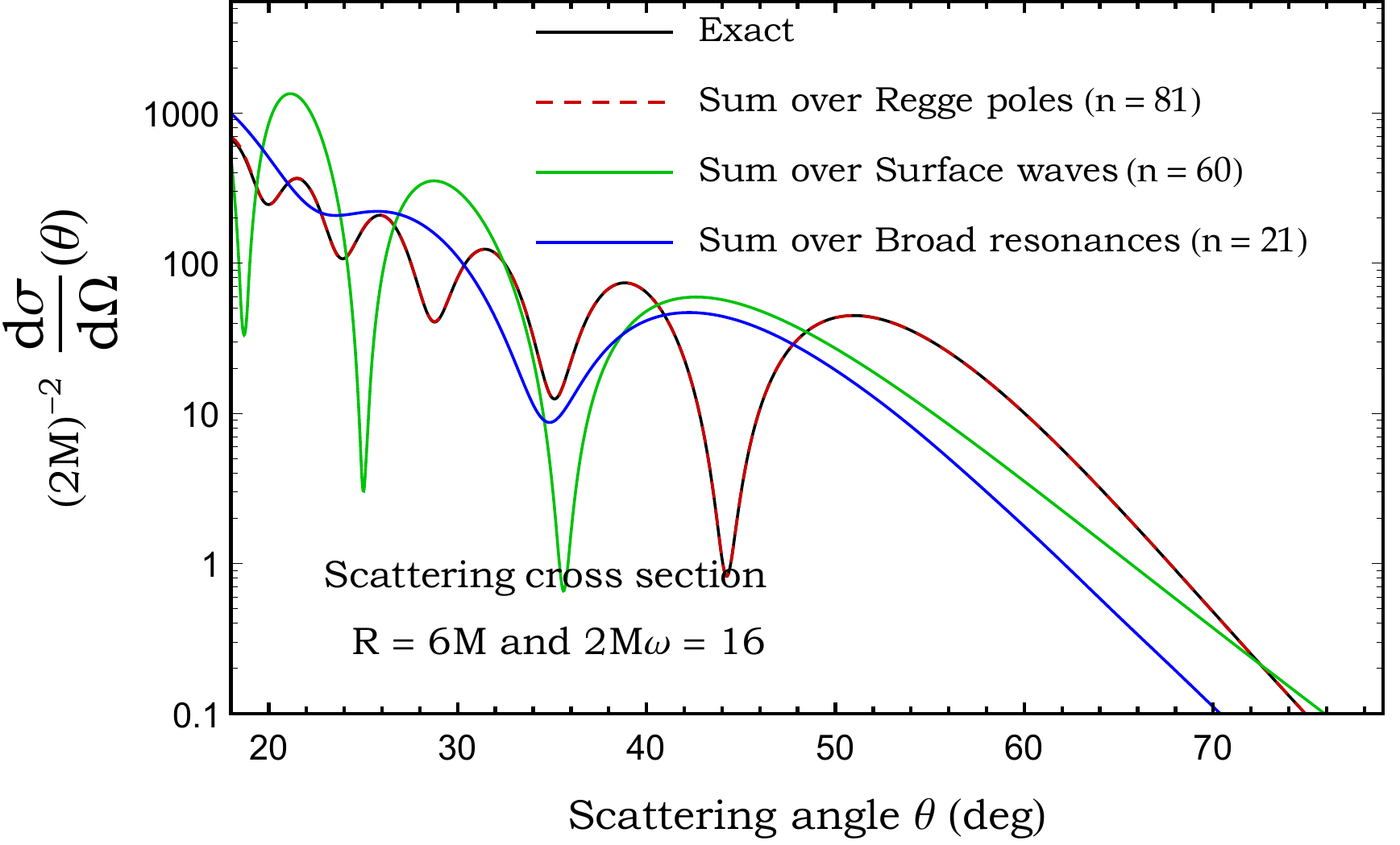}
  \vspace*{-0.45cm}
\caption{\label{Rainbow_Cross_Section_R_6_2Mw_16} Rainbow scattering for compact bodies for $2M\omega=16$ and $R=6M$, its Regge pole approximation and different contributions of the sum over Regge poles.}
\end{figure}

\begin{figure}[htp!]
\centering
 \includegraphics[scale=0.50]{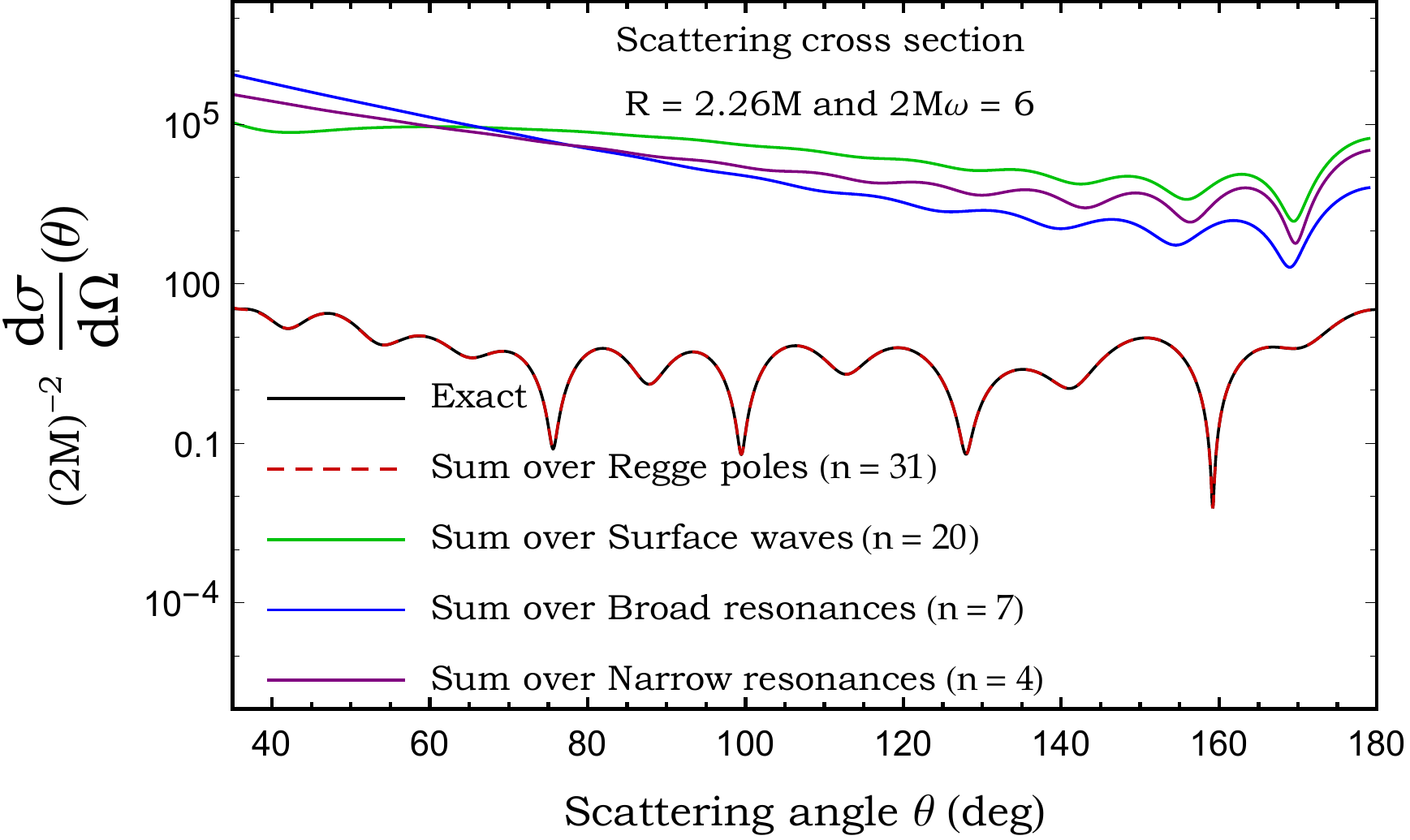}
  \vspace*{-0.45cm}
\caption{\label{Diff_Contribution_Cross_Section_R_2-dot-26_2Mw_6} Rainbow scattering for UCOs for $2M\omega=16$ and $R=2.26M$, its Regge pole approximation and different contributions of the sum over Regge poles.}
\end{figure}

Figures \ref{S_0_R_6_2Mw_3_Exact_vs_CAM}, \ref{S_0_R_6_2Mw_16_Exact_vs_CAM}, \ref{S_0_R_2-dot-26_2Mw_3_Exact_vs_CAM}, \ref{S_0_R_2-dot-26_2Mw_6_Exact_vs_CAM}, \ref{Rainbow_Cross_Section_R_6_2Mw_16}, and \ref{Diff_Contribution_Cross_Section_R_2-dot-26_2Mw_6} show a selection of results for the scattering cross section $d\sigma / d\Omega$ computed with the CAM method, and compared with results from the standard partial-wave method. As in Sec.~\ref{subsec:results1}, we focus on two cases: a neutron-starlike object with $R/M = 6$, and an ultracompact object with $R/M = 2.26$.

Figure \ref{S_0_R_6_2Mw_3_Exact_vs_CAM} demonstrates that the CAM cross section approaches the partial-wave cross section as (i) successively more Regge poles are included in the sum (\ref{CAM_Scalar_Scattering_amp_decomp_RP}), and (ii) the background-integral contributions (\ref{CAM_Scalar_Scattering_amp_decomp_Background}) are included. The first plot shows that, at a ``low'' frequency of $M \omega = 3/2$ for the neutron star model ($R/M = 6$), taking a sum over just \emph{two} Regge poles in Eq.~(\ref{CAM_Scalar_Scattering_amp_decomp_RP}) captures the crude features of the cross section. The final plot shows that a cross section calculated from a sum of $15$ Regge poles and the background integral is indistinguishable (on the plot) from the partial-wave sum.

Figure \ref{S_0_R_6_2Mw_16_Exact_vs_CAM} shows the neutron star model ($R/M = 6$) at the higher frequency of $M \omega = 8$. The first plot shows that, in this case, including just two Regge poles is yields a poor approximation. With 20 Regge poles, the primary peak of the rainbow, the first supernumerary, and the shadow region are well captured, but at smaller angles ($\theta  \lesssim 40^\circ$) there is no agreement. With 42 Regge poles, but no background integral, the agreement is excellent for $\theta \gtrsim 30^\circ$. Finally, with 81 Regge poles and the background integral included, the CAM result is again indistinguishable from the partial-wave result on the plot.

Figure \ref{S_0_R_2-dot-26_2Mw_3_Exact_vs_CAM} shows the cross section for an ultracompact object with $R/M = 2.26$ at the ``low'' frequency of $M \omega = 3/2$. In this case, the rainbow angle exceeds $180^\circ$, and there is both a light-ring and a trapping region (see Fig.~\ref{fig:Veff}). The cross section exhibits regular orbiting oscillations with angle $\theta$. The angular width is consistent with generation by the light-ring (i.e.~ the peak in the potential barrier). The second plot shows that including just 3 modes from each of the three branches leads to a good description of scattering at large angles $\theta \gtrsim 100^\circ$. As there is no shadow region in this case, the cross section at large angles in non-negligible. For $R/M = 2.26$, the cross section at the antipodal point ($\theta = 180^\circ$) is a factor of $> 10^6$ larger than in the $R/M = 6$ case. The final plot shows that excellent agreement with the partial-wave results is obtained, for $\theta \gtrsim 20^\circ$, by summing over $27$ poles.

Figure \ref{S_0_R_2-dot-26_2Mw_6_Exact_vs_CAM} shows the cross section for an UCO at the higher frequency of $M \omega = 8$. In this case, the cross section is non-negligible at all angles, and there is evidence for interference between oscillations of comparable amplitudes and widths. Summing 31 Regge poles leads to good agreement for the cross section for angles $\theta \gtrsim 35^\circ$. As before, it is necessary to include the background integral as well to obtain agreement at smaller angles.

Figures \ref{Rainbow_Cross_Section_R_6_2Mw_16} and \ref{Diff_Contribution_Cross_Section_R_2-dot-26_2Mw_6} show the relative magnitudes of the contributions from the different branches of Regge poles. For $R/M = 6$ (Fig.~\ref{Rainbow_Cross_Section_R_6_2Mw_16}) the surface-wave and broad-resonance amplitudes are similar in magnitude, and similar in magnitude to their sum. There is a difference in phase between these amplitudes which generates the peaks and troughs of the rainbow-scattering pattern.

For $R/M = 2.26$ (Fig.~\ref{Diff_Contribution_Cross_Section_R_2-dot-26_2Mw_6}), however, the amplitudes  from the three branches are comparable in magnitude, but their sum is several orders-of-magnitude smaller. In other words, there is some delicate cancellation occurring between the contributions from the branches, suggesting that the CAM approach used here is not an efficient method for computing the cross section in this case.

\section{Discussion and conclusions\label{sec:conclusions}}

In this work, we have: (1) calculated the full spectrum of Regge poles for a compact body spacetime for the first time; (2) applied CAM theory to obtain an expression for the scattering amplitude as a residue series accompanied by a background integral, Eqs.~(\ref{CAM_Scalar_Scattering_amp_tot})--(\ref{CAM_Scalar_Scattering_amp_decomp_RP}); and (3) utilised the CAM formulas to calculate scattering cross sections numerically, demonstrating precise agreement with the numerical results of the partial-wave expansion, first computed in Ref.~\cite{Dolan:2017rtj}.

The spectrum of Regge poles exhibits two distinct branches of poles in the neutron star case $R/M = 6$ (see Fig.~\ref{RP_R6_approx_2Mw_3_6_s_1}), which we have labeled ``broad resonances'' and ``surface waves''. Ultracompact objects have, in addition, a third branch of ``narrow resonances'' (see Fig.~\ref{RP_R226_approx_2Mw_3_6_s_1}), linked to the existence of an effective cavity between the light-ring and the object's surface. Similar branches of poles arise in scattering by a transparent sphere, and so it is natural to adopt the terminology used to describe these branches in Mie scattering \cite{Nussenzveig:2006}.

By applying a WKB method at lowest order, we have obtained a better understanding of the spectrum of the broad resonances, finding that the magnitude of the imaginary part is linked to the magnitude of the discontinuity in the effective potential at the surface of the body (see Eq.~(\ref{eq:alpha}), Eq.~(\ref{lambda_Approx}) and Fig.~\ref{fig:Veff}). Extending the WKB method to higher orders would improve the accuracy of this approximation, and further work is needed to obtain asymptotic approximations for the surface-wave branch, and the narrow-resonance branch.

We can be confident that there are no additional branches of Regge poles in the first quadrant, for two reasons. First, we have scanned across the complex-$\lambda$ domains shown in Figs.~\ref{RP_R6_approx_2Mw_3_6_s_1} and \ref{RP_R226_approx_2Mw_3_6_s_1}, as detailed in Sec.~\ref{subsec:method}. Second, in our calculation of $d\sigma/d\Omega$, we find a precise numerical agreement between the CAM calculation and the partial-wave calculation; this would not be the case if poles in this domain had been missed.

The CAM toolkit was applied here to gain a complementary method for calculating scattering amplitudes. In summary, we have shown that, for the intermediate and high angles, these scattering amplitudes can be reconstructed in terms of Regge poles with great precision in the intermediate and high frequency regimes. For compact objects ($R/M\sim6$), we observed that it is necessary to take into account the background integral contributions to describe the glory, and this is true regardless of the frequency. Conversely, this is not the case for UCOs ($R/M < 3$), where the sum over the Regge poles is sufficient to accurately describe the glory. Moreover, the sum over the Regge poles permitted us to overcome the difficulties linked to the lack of convergence from which the scattering amplitude suffers (\textit{i.e.} partial wave expansions) due to the long-range of the field propagating on the Schwarzschild spacetime outside the compact body. However, the existence of the broad resonance branch 
close to the real axis with an approximately uniform spacing, $\Delta\lambda\sim 2$, requires the calculation of a large number of poles. In the black hole case, by contrast, there is only a single branch of surface-wave-type modes (see Fig.~\ref{RP_R6_approx_2Mw_3_6_s_1}), and in the high-frequency limit, a few poles ($<5$) capture both the orbiting and glory phenomena \cite{Folacci:2019cmc, Folacci:2019vtt}.

Here we have numerically determined if a background integral, and how many Regge-poles and residues, need to be calculated to accurately reproduce a scattering cross section in an angular region. Loosely speaking, for scattering at higher frequencies, more poles must be included since there are more broad resonances. Also the background integral is negligible except for weak glory scattering. This is an interesting result in itself (as noted in \cite{Folacci:2019cmc,Folacci:2019vtt}) since in most CAM scattering studies the background integral is significant in geometrically illuminated regions \cite{Nussenzveig:2006}.
An asymptotic analysis is an important next step toward understanding the role of the background integral and higher Regge-pole overtones, both computationally and, in a semiclassical interpretation, physically.

In the case of Mie scattering, a powerful extension of the CAM approach is to define the so-called Regge-Debye poles using the Debye expansion (see the chapter 9 of the monograph of Nussenzveig \cite{Nussenzveig:2006}), thereby eliminating the broad resonance branch entirely, and accelerating the convergence. This is another possible direction for a future investigation.


\acknowledgments
M.O.E.H. wishes to thank Antoine Folacci for various discussions concerning this work and thanks S.R.D.~for his kind
invitation to the University of Sheffield. T.S.~acknowledges financial support from EPSRC. S.R.D.~acknowledges financial support from the European Union's Horizon 2020 research and innovation programme under the H2020-MSCA-RISE-2017 Grant No.~FunFiCO-777740, and from the Science and Technology Facilities Council (STFC) under Grant No.~ST/P000800/1.

\appendix

\section{Location of the Regge poles in complex $\lambda$-plan}
\label{appen}



In this appendix, we establish which quadrants of the complex $\lambda$-plane may contain Regge poles, for scalar and axial metric perturbations on a stellar background. In other words, we will constrain the possible locations of the zeros of the coefficient $A^{(-)}_{\lambda-1/2} (\omega)$ [see Eq.~\eqref{Matrix_S_CAM}],
\begin{equation}\label{zeros_Am}
   A^{(-)}_{\lambda-1/2} (\omega)\Big{|}_{\lambda =\lambda_n} = 0,
\end{equation}
for $\omega$ real.

To do this, it is convenient to write the boundary conditions \eqref{bc_1_in} and \eqref{bc_2_in} defining a solution that is regular at the origin in the following form:
\begin{equation}
\label{bc_CAM}
\phi_{\omega,\lambda-\frac{1}{2}}(r) \sim
\begin{cases}
r^{\lambda+\frac{1}{2}}, &
r \rightarrow 0, \\
\phantom{+} A_{\lambda-1/2}^{(-)}(\omega) e^{-i\omega r_*} & \\
 + A_{\lambda-1/2}^{(+)}(\omega) e^{+i\omega r_*} , &r_* \rightarrow +\infty.
\end{cases}
\end{equation}
We next consider the radial equation~\eqref{H_Radial_equation} for the regular solution
\begin{equation}\label{Wave_equation_reg}
   \frac{d^2}{dr_*^2}\phi_{\omega,\lambda-1/2}(r)+\left(\omega^2 -V_{\lambda-1/2}\right) \phi_{\omega,\lambda-1/2}(r) =0,
 \end{equation}
and its complex conjugate,
 \begin{equation}\label{Wave_equation_reg_conjugate}
   \frac{d^2}{dr_*^2}\phi^*_{\omega,\lambda-1/2}(r)+ \left(\omega^2 -V_{\lambda^*-1/2}\right)\phi^*_{\omega,{\lambda-1/2}}(r) = 0 .
 \end{equation}
Multiplying \eqref{Wave_equation_reg} by $\phi^*_{\omega,\lambda-1/2}(r)$ and \eqref{Wave_equation_reg_conjugate} by $\phi_{\omega,\lambda-1/2}(r)$ and taking the difference, it follows that
\begin{widetext}
\begin{equation}\label{Derivative_Wronskian}
\begin{aligned}
 \phi^*_{\omega,\lambda-1/2}(r) \frac{d^2}{dr_*^2}\phi_{\omega,\lambda-1/2}(r) - \phi_{\omega,\lambda-1/2}(r) \frac{d^2}{dr_*^2} \phi^*_{\omega,\lambda-1/2}(r) 
                           &= 2 i|\phi_{\lambda-1/2}(\omega)|^2 f[r]\frac{\text{Im}[\lambda^2]}{r^2} .
 \end{aligned}
\end{equation}
\end{widetext}
The LHS of Eq.~\eqref{Derivative_Wronskian} is the derivative of $W$, the Wronskian of the two solutions with respect to $r_*$. The Wronskian is a constant (\textit{i.e.}, $d/dr_* W[\phi^*,\phi] = 0$) if $\lambda$ either real or purely imaginary. According to Eq.~\eqref{Derivative_Wronskian}, we can write
\begin{equation}\label{Derivative_Wronskian_bis}
\frac{d}{dr_*}W[\phi^*,\phi]= 2i|\phi_{\lambda-1/2}(\omega)|^2 f[r]\frac{2\text{Re}[\lambda]\text{Im}[\lambda]}{r^2}.
\end{equation}

Integration of \eqref{Derivative_Wronskian_bis} gives
\begin{equation}\label{Evaluation_W}
\lim_{r_*\rightarrow \infty}  W[\phi^*,\phi] = 4i\, \text{Im}[\lambda]\, \text{Re}[\lambda] \int_{0}^{+\infty} \frac{|\phi_{\lambda-1/2}(\omega)|^2}{r^2}\,dr ,
\end{equation}
where we have used $\frac{dr}{dr_*}=f(r)$ to transform the integration variable. On the other hand,  by evaluating the  Wronskian from the boundary condition~\eqref{bc_CAM}, we obtain
\begin{equation}\label{Evaluation_W_bis}
\lim_{r_*\rightarrow \infty}  W[\phi^*,\phi] =-2i\, \omega\left(|A^{(-)}_{\lambda-1/2}|^2-|A^{(+)}_{\lambda-1/2}|^2\right),
\end{equation}
hence we get (\textit{c.f.}~\eqref{Evaluation_W} and  \eqref{Evaluation_W_bis})
\begin{equation}\label{Evaluation_W_bis_1}
  |A^{(-)}_{\lambda-1/2}|^2-|A^{(+)}_{\lambda-1/2}|^2 = -\frac{2}{\omega}\text{Re}[\lambda]\text{Im}[\lambda]\int_{0}^{+\infty} \frac{ |\phi_{\lambda-1/2}(\omega)|^2}{r^2}\,dr.
\end{equation}

Since the integral is positive and $\omega$ is real, $A^{(-)}_{\lambda-1/2}$ can  vanish in the right-half complex $\lambda$-plan (\text{i.e.} $\text{Re}[\lambda] >0 $) only for
\begin{equation}
    \begin{dcases}
        \omega>0 \\
        \text{Im}[\lambda]>0
     \end{dcases}
\quad\text{or}\quad
    \begin{dcases}
      \omega<0 \\
        \text{Im}[\lambda]<0
    \end{dcases}
\end{equation}\

Thus, in the right-half complex $\lambda$-plan, the Regge poles lie only in the first quadrant for $\omega > 0$, and only in the fourth quadrant for $\omega < 0$.

It is important to note that under the transformation $\omega \rightarrow -\omega$, the coefficient $A^{(-)}_{\lambda-1/2}(\omega)$ defined by \eqref{H_Radial_equation}--\eqref{bc_1_in}--\eqref{bc_2_in} satisfies the symmetry relation
\begin{equation}\label{symmetry_relation}
 A^{(-)}_{\lambda-1/2}(-\omega) = \left[A^{(-)}_{\lambda^*-1/2}(\omega)\right]^*,
\end{equation}
and thus
\begin{equation}\label{Relation_1st_4th}
\lambda_n(-\omega) = \lambda_n(\omega)^*.
\end{equation}


As discussed in Sec.~\ref{subsec:results1}, a good analytic extension of the S-matrix into the left-half complex $\lambda$-plane leads to the symmetry \eqref{Matrix_S_CAM_symm}. Then, it follows that the Regge poles with $\text{Re}[\lambda] < 0 $ may be found for (\textit{c.f.} \eqref{Evaluation_W_bis_1} and \eqref{Relation_1st_4th})
\begin{equation}
    \begin{dcases}
        \omega>0 \\
        \text{Im}[\lambda]<0
     \end{dcases}
\quad\text{or}\quad
    \begin{dcases}
      \omega<0 \\
        \text{Im}[\lambda]>0
    \end{dcases}.
\end{equation}\

The results established here agree with the situation in non-relativistic wave and atom/particle scattering (see \textit{e.g.}~\cite{Newton:1982qc,Bottino1962} and references therein), and also with the results of Schwarzschild black hole scattering \cite{Andersson:1994rk,Decanini:2009mu}.

\bibliography{rainbowCAM}

\begin{thebibliography}{78}%
\makeatletter
\providecommand \@ifxundefined [1]{%
 \@ifx{#1\undefined}
}%
\providecommand \@ifnum [1]{%
 \ifnum #1\expandafter \@firstoftwo
 \else \expandafter \@secondoftwo
 \fi
}%
\providecommand \@ifx [1]{%
 \ifx #1\expandafter \@firstoftwo
 \else \expandafter \@secondoftwo
 \fi
}%
\providecommand \natexlab [1]{#1}%
\providecommand \enquote  [1]{``#1''}%
\providecommand \bibnamefont  [1]{#1}%
\providecommand \bibfnamefont [1]{#1}%
\providecommand \citenamefont [1]{#1}%
\providecommand \href@noop [0]{\@secondoftwo}%
\providecommand \href [0]{\begingroup \@sanitize@url \@href}%
\providecommand \@href[1]{\@@startlink{#1}\@@href}%
\providecommand \@@href[1]{\endgroup#1\@@endlink}%
\providecommand \@sanitize@url [0]{\catcode `\\12\catcode `\$12\catcode
  `\&12\catcode `\#12\catcode `\^12\catcode `\_12\catcode `\%12\relax}%
\providecommand \@@startlink[1]{}%
\providecommand \@@endlink[0]{}%
\providecommand \url  [0]{\begingroup\@sanitize@url \@url }%
\providecommand \@url [1]{\endgroup\@href {#1}{\urlprefix }}%
\providecommand \urlprefix  [0]{URL }%
\providecommand \Eprint [0]{\href }%
\providecommand \doibase [0]{http://dx.doi.org/}%
\providecommand \selectlanguage [0]{\@gobble}%
\providecommand \bibinfo  [0]{\@secondoftwo}%
\providecommand \bibfield  [0]{\@secondoftwo}%
\providecommand \translation [1]{[#1]}%
\providecommand \BibitemOpen [0]{}%
\providecommand \bibitemStop [0]{}%
\providecommand \bibitemNoStop [0]{.\EOS\space}%
\providecommand \EOS [0]{\spacefactor3000\relax}%
\providecommand \BibitemShut  [1]{\csname bibitem#1\endcsname}%
\let\auto@bib@innerbib\@empty
\bibitem [{\citenamefont {{Hildreth}}(1964)}]{Hildreth1964PhDT64}%
  \BibitemOpen
  \bibfield  {author} {\bibinfo {author} {\bibfnamefont {W.~W.}\ \bibnamefont
  {{Hildreth}}},\ }\emph {\bibinfo {title} {{The Interaction of Scalar
  Gravitational Waves with the Schwarzschild Metric.}}},\ \href@noop {} {Ph.D.
  thesis},\ \bibinfo  {school} {Princeton University} (\bibinfo {year}
  {1964})\BibitemShut {NoStop}%
\bibitem [{\citenamefont {Matzner}(1968)}]{Matzner:1968}%
  \BibitemOpen
  \bibfield  {author} {\bibinfo {author} {\bibfnamefont {R.~A.}\ \bibnamefont
  {Matzner}},\ }\href@noop {} {\bibfield  {journal} {\bibinfo  {journal}
  {Journal of Mathematical Physics}\ }\textbf {\bibinfo {volume} {9}},\
  \bibinfo {pages} {163} (\bibinfo {year} {1968})}\BibitemShut {NoStop}%
\bibitem [{\citenamefont {Vishveshwara}(1970)}]{Vishveshwara:1970}%
  \BibitemOpen
  \bibfield  {author} {\bibinfo {author} {\bibfnamefont {C.}~\bibnamefont
  {Vishveshwara}},\ }\href@noop {} {\bibfield  {journal} {\bibinfo  {journal}
  {Nature}\ }\textbf {\bibinfo {volume} {227}},\ \bibinfo {pages} {936}
  (\bibinfo {year} {1970})}\BibitemShut {NoStop}%
\bibitem [{\citenamefont {Chrzanowski}\ \emph {et~al.}(1976)\citenamefont
  {Chrzanowski}, \citenamefont {Matzner}, \citenamefont {Sandberg},\ and\
  \citenamefont {Ryan}}]{Chrzanowski:1976jb}%
  \BibitemOpen
  \bibfield  {author} {\bibinfo {author} {\bibfnamefont {P.~L.}\ \bibnamefont
  {Chrzanowski}}, \bibinfo {author} {\bibfnamefont {R.~A.}\ \bibnamefont
  {Matzner}}, \bibinfo {author} {\bibfnamefont {V.~D.}\ \bibnamefont
  {Sandberg}}, \ and\ \bibinfo {author} {\bibfnamefont {M.~P.}\ \bibnamefont
  {Ryan}},\ }\href {\doibase 10.1103/PhysRevD.14.317} {\bibfield  {journal}
  {\bibinfo  {journal} {Phys. Rev.}\ }\textbf {\bibinfo {volume} {D14}},\
  \bibinfo {pages} {317} (\bibinfo {year} {1976})}\BibitemShut {NoStop}%
\bibitem [{\citenamefont {Mashhoon}(1973)}]{Mashhoon:1973zz}%
  \BibitemOpen
  \bibfield  {author} {\bibinfo {author} {\bibfnamefont {B.}~\bibnamefont
  {Mashhoon}},\ }\href {\doibase 10.1103/PhysRevD.7.2807} {\bibfield  {journal}
  {\bibinfo  {journal} {Phys. Rev.}\ }\textbf {\bibinfo {volume} {D7}},\
  \bibinfo {pages} {2807} (\bibinfo {year} {1973})}\BibitemShut {NoStop}%
\bibitem [{\citenamefont {Fabbri}(1975)}]{Fabbri:1975}%
  \BibitemOpen
  \bibfield  {author} {\bibinfo {author} {\bibfnamefont {R.}~\bibnamefont
  {Fabbri}},\ }\href@noop {} {\bibfield  {journal} {\bibinfo  {journal}
  {Physical Review D}\ }\textbf {\bibinfo {volume} {12}},\ \bibinfo {pages}
  {933} (\bibinfo {year} {1975})}\BibitemShut {NoStop}%
\bibitem [{\citenamefont {Sanchez}(1978)}]{Sanchez:1977vz}%
  \BibitemOpen
  \bibfield  {author} {\bibinfo {author} {\bibfnamefont {N.~G.}\ \bibnamefont
  {Sanchez}},\ }\href {\doibase 10.1103/PhysRevD.18.1798} {\bibfield  {journal}
  {\bibinfo  {journal} {Phys. Rev.}\ }\textbf {\bibinfo {volume} {D18}},\
  \bibinfo {pages} {1798} (\bibinfo {year} {1978})}\BibitemShut {NoStop}%
\bibitem [{\citenamefont {Matzner}\ and\ \citenamefont
  {Ryan}(1978)}]{MatznerRyan1978}%
  \BibitemOpen
  \bibfield  {author} {\bibinfo {author} {\bibfnamefont {R.~A.}\ \bibnamefont
  {Matzner}}\ and\ \bibinfo {author} {\bibfnamefont {M.~P.~J.}\ \bibnamefont
  {Ryan}},\ }\href {\doibase 10.1086/190508} {\bibfield  {journal} {\bibinfo
  {journal} {The Astrophysical Journal Supplement Series}\ }\textbf {\bibinfo
  {volume} {36}},\ \bibinfo {pages} {451} (\bibinfo {year} {1978})}\BibitemShut
  {NoStop}%
\bibitem [{\citenamefont {Handler}\ and\ \citenamefont
  {Matzner}(1980)}]{Handler:1980un}%
  \BibitemOpen
  \bibfield  {author} {\bibinfo {author} {\bibfnamefont {F.~A.}\ \bibnamefont
  {Handler}}\ and\ \bibinfo {author} {\bibfnamefont {R.~A.}\ \bibnamefont
  {Matzner}},\ }\href {\doibase 10.1103/PhysRevD.22.2331} {\bibfield  {journal}
  {\bibinfo  {journal} {Phys. Rev.}\ }\textbf {\bibinfo {volume} {D22}},\
  \bibinfo {pages} {2331} (\bibinfo {year} {1980})}\BibitemShut {NoStop}%
\bibitem [{\citenamefont {Matzner}\ \emph {et~al.}(1985)\citenamefont
  {Matzner}, \citenamefont {DeWitt-Morette}, \citenamefont {Nelson},\ and\
  \citenamefont {Zhang}}]{Matzner:1985rjn}%
  \BibitemOpen
  \bibfield  {author} {\bibinfo {author} {\bibfnamefont {R.~A.}\ \bibnamefont
  {Matzner}}, \bibinfo {author} {\bibfnamefont {C.}~\bibnamefont
  {DeWitt-Morette}}, \bibinfo {author} {\bibfnamefont {B.}~\bibnamefont
  {Nelson}}, \ and\ \bibinfo {author} {\bibfnamefont {T.-R.}\ \bibnamefont
  {Zhang}},\ }\href {\doibase 10.1103/PhysRevD.31.1869} {\bibfield  {journal}
  {\bibinfo  {journal} {Phys. Rev.}\ }\textbf {\bibinfo {volume} {D31}},\
  \bibinfo {pages} {1869} (\bibinfo {year} {1985})}\BibitemShut {NoStop}%
\bibitem [{\citenamefont {Futterman}\ \emph {et~al.}(2012)\citenamefont
  {Futterman}, \citenamefont {Handler},\ and\ \citenamefont
  {Matzner}}]{Futterman:1988ni}%
  \BibitemOpen
  \bibfield  {author} {\bibinfo {author} {\bibfnamefont {J.~A.~H.}\
  \bibnamefont {Futterman}}, \bibinfo {author} {\bibfnamefont {F.~A.}\
  \bibnamefont {Handler}}, \ and\ \bibinfo {author} {\bibfnamefont {R.~A.}\
  \bibnamefont {Matzner}},\ }\href@noop {} {\emph {\bibinfo {title}
  {{Scattering from black holes}}}}\ (\bibinfo  {publisher} {Cambridge
  University Press},\ \bibinfo {year} {2012})\BibitemShut {NoStop}%
\bibitem [{\citenamefont {Andersson}(1995)}]{Andersson:1995vi}%
  \BibitemOpen
  \bibfield  {author} {\bibinfo {author} {\bibfnamefont {N.}~\bibnamefont
  {Andersson}},\ }\href {\doibase 10.1103/PhysRevD.52.1808} {\bibfield
  {journal} {\bibinfo  {journal} {Phys. Rev.}\ }\textbf {\bibinfo {volume}
  {D52}},\ \bibinfo {pages} {1808} (\bibinfo {year} {1995})}\BibitemShut
  {NoStop}%
\bibitem [{\citenamefont {Glampedakis}\ and\ \citenamefont
  {Andersson}(2001)}]{Glampedakis:2001cx}%
  \BibitemOpen
  \bibfield  {author} {\bibinfo {author} {\bibfnamefont {K.}~\bibnamefont
  {Glampedakis}}\ and\ \bibinfo {author} {\bibfnamefont {N.}~\bibnamefont
  {Andersson}},\ }\href {\doibase 10.1088/0264-9381/18/10/309} {\bibfield
  {journal} {\bibinfo  {journal} {Class. Quant. Grav.}\ }\textbf {\bibinfo
  {volume} {18}},\ \bibinfo {pages} {1939} (\bibinfo {year} {2001})},\ \Eprint
  {http://arxiv.org/abs/gr-qc/0102100} {arXiv:gr-qc/0102100 [gr-qc]}
  \BibitemShut {NoStop}%
\bibitem [{\citenamefont {Dolan}\ \emph {et~al.}(2006)\citenamefont {Dolan},
  \citenamefont {Doran},\ and\ \citenamefont {Lasenby}}]{Dolan:2006vj}%
  \BibitemOpen
  \bibfield  {author} {\bibinfo {author} {\bibfnamefont {S.}~\bibnamefont
  {Dolan}}, \bibinfo {author} {\bibfnamefont {C.}~\bibnamefont {Doran}}, \ and\
  \bibinfo {author} {\bibfnamefont {A.}~\bibnamefont {Lasenby}},\ }\href
  {\doibase 10.1103/PhysRevD.74.064005} {\bibfield  {journal} {\bibinfo
  {journal} {Phys. Rev.}\ }\textbf {\bibinfo {volume} {D74}},\ \bibinfo {pages}
  {064005} (\bibinfo {year} {2006})},\ \Eprint
  {http://arxiv.org/abs/gr-qc/0605031} {arXiv:gr-qc/0605031 [gr-qc]}
  \BibitemShut {NoStop}%
\bibitem [{\citenamefont {Dolan}(2008{\natexlab{a}})}]{Dolan:2007ut}%
  \BibitemOpen
  \bibfield  {author} {\bibinfo {author} {\bibfnamefont {S.~R.}\ \bibnamefont
  {Dolan}},\ }\href {\doibase 10.1103/PhysRevD.77.044004} {\bibfield  {journal}
  {\bibinfo  {journal} {Phys. Rev.}\ }\textbf {\bibinfo {volume} {D77}},\
  \bibinfo {pages} {044004} (\bibinfo {year} {2008}{\natexlab{a}})},\ \Eprint
  {http://arxiv.org/abs/0710.4252} {arXiv:0710.4252 [gr-qc]} \BibitemShut
  {NoStop}%
\bibitem [{\citenamefont {Dolan}(2008{\natexlab{b}})}]{Dolan:2008kf}%
  \BibitemOpen
  \bibfield  {author} {\bibinfo {author} {\bibfnamefont {S.~R.}\ \bibnamefont
  {Dolan}},\ }\href {\doibase 10.1088/0264-9381/25/23/235002} {\bibfield
  {journal} {\bibinfo  {journal} {Class. Quant. Grav.}\ }\textbf {\bibinfo
  {volume} {25}},\ \bibinfo {pages} {235002} (\bibinfo {year}
  {2008}{\natexlab{b}})},\ \Eprint {http://arxiv.org/abs/0801.3805}
  {arXiv:0801.3805 [gr-qc]} \BibitemShut {NoStop}%
\bibitem [{\citenamefont {Crispino}\ \emph {et~al.}(2009)\citenamefont
  {Crispino}, \citenamefont {Dolan},\ and\ \citenamefont
  {Oliveira}}]{Crispino:2009xt}%
  \BibitemOpen
  \bibfield  {author} {\bibinfo {author} {\bibfnamefont {L.~C.~B.}\
  \bibnamefont {Crispino}}, \bibinfo {author} {\bibfnamefont {S.~R.}\
  \bibnamefont {Dolan}}, \ and\ \bibinfo {author} {\bibfnamefont {E.~S.}\
  \bibnamefont {Oliveira}},\ }\href {\doibase 10.1103/PhysRevLett.102.231103}
  {\bibfield  {journal} {\bibinfo  {journal} {Phys. Rev. Lett.}\ }\textbf
  {\bibinfo {volume} {102}},\ \bibinfo {pages} {231103} (\bibinfo {year}
  {2009})},\ \Eprint {http://arxiv.org/abs/0905.3339} {arXiv:0905.3339 [gr-qc]}
  \BibitemShut {NoStop}%
\bibitem [{\citenamefont {Cotaescu}\ \emph {et~al.}(2016)\citenamefont
  {Cotaescu}, \citenamefont {Crucean},\ and\ \citenamefont
  {Sporea}}]{Cotaescu:2014jca}%
  \BibitemOpen
  \bibfield  {author} {\bibinfo {author} {\bibfnamefont {I.~I.}\ \bibnamefont
  {Cotaescu}}, \bibinfo {author} {\bibfnamefont {C.}~\bibnamefont {Crucean}}, \
  and\ \bibinfo {author} {\bibfnamefont {C.~A.}\ \bibnamefont {Sporea}},\
  }\href {\doibase 10.1140/epjc/s10052-016-3936-9} {\bibfield  {journal}
  {\bibinfo  {journal} {Eur. Phys. J.}\ }\textbf {\bibinfo {volume} {C76}},\
  \bibinfo {pages} {102} (\bibinfo {year} {2016})},\ \Eprint
  {http://arxiv.org/abs/1409.7201} {arXiv:1409.7201 [gr-qc]} \BibitemShut
  {NoStop}%
\bibitem [{\citenamefont {Gu{\ss}mann}(2017)}]{Gussmann:2016mkp}%
  \BibitemOpen
  \bibfield  {author} {\bibinfo {author} {\bibfnamefont {A.}~\bibnamefont
  {Gu{\ss}mann}},\ }\href {\doibase 10.1088/1361-6382/aa606c} {\bibfield
  {journal} {\bibinfo  {journal} {Class. Quant. Grav.}\ }\textbf {\bibinfo
  {volume} {34}},\ \bibinfo {pages} {065007} (\bibinfo {year} {2017})},\
  \Eprint {http://arxiv.org/abs/1608.00552} {arXiv:1608.00552 [hep-th]}
  \BibitemShut {NoStop}%
\bibitem [{\citenamefont {De~Logi}\ and\ \citenamefont
  {Kovacs}(1977)}]{DeLogi:1977dp}%
  \BibitemOpen
  \bibfield  {author} {\bibinfo {author} {\bibfnamefont {W.~K.}\ \bibnamefont
  {De~Logi}}\ and\ \bibinfo {author} {\bibfnamefont {S.~J.}\ \bibnamefont
  {Kovacs}},\ }\href {\doibase 10.1103/PhysRevD.16.237} {\bibfield  {journal}
  {\bibinfo  {journal} {Phys. Rev.}\ }\textbf {\bibinfo {volume} {D16}},\
  \bibinfo {pages} {237} (\bibinfo {year} {1977})}\BibitemShut {NoStop}%
\bibitem [{\citenamefont {Guadagnini}(2008)}]{Guadagnini:2008ha}%
  \BibitemOpen
  \bibfield  {author} {\bibinfo {author} {\bibfnamefont {E.}~\bibnamefont
  {Guadagnini}},\ }\href {\doibase 10.1088/0264-9381/25/9/095012} {\bibfield
  {journal} {\bibinfo  {journal} {Class. Quant. Grav.}\ }\textbf {\bibinfo
  {volume} {25}},\ \bibinfo {pages} {095012} (\bibinfo {year} {2008})},\
  \Eprint {http://arxiv.org/abs/0803.2855} {arXiv:0803.2855 [gr-qc]}
  \BibitemShut {NoStop}%
\bibitem [{\citenamefont {Sorge}(2015)}]{Sorge:2015yoa}%
  \BibitemOpen
  \bibfield  {author} {\bibinfo {author} {\bibfnamefont {F.}~\bibnamefont
  {Sorge}},\ }\href {\doibase 10.1088/0264-9381/32/3/035007} {\bibfield
  {journal} {\bibinfo  {journal} {Class. Quant. Grav.}\ }\textbf {\bibinfo
  {volume} {32}},\ \bibinfo {pages} {035007} (\bibinfo {year}
  {2015})}\BibitemShut {NoStop}%
\bibitem [{\citenamefont {Anninos}\ \emph {et~al.}(1992)\citenamefont
  {Anninos}, \citenamefont {DeWitt-Morette}, \citenamefont {Matzner},
  \citenamefont {Yioutas},\ and\ \citenamefont {Zhang}}]{Anninos:1992ih}%
  \BibitemOpen
  \bibfield  {author} {\bibinfo {author} {\bibfnamefont {P.}~\bibnamefont
  {Anninos}}, \bibinfo {author} {\bibfnamefont {C.}~\bibnamefont
  {DeWitt-Morette}}, \bibinfo {author} {\bibfnamefont {R.~A.}\ \bibnamefont
  {Matzner}}, \bibinfo {author} {\bibfnamefont {P.}~\bibnamefont {Yioutas}}, \
  and\ \bibinfo {author} {\bibfnamefont {T.~R.}\ \bibnamefont {Zhang}},\ }\href
  {\doibase 10.1103/PhysRevD.46.4477} {\bibfield  {journal} {\bibinfo
  {journal} {Phys. Rev.}\ }\textbf {\bibinfo {volume} {D46}},\ \bibinfo {pages}
  {4477} (\bibinfo {year} {1992})}\BibitemShut {NoStop}%
\bibitem [{\citenamefont {Leite}\ \emph {et~al.}(2019)\citenamefont {Leite},
  \citenamefont {Benone},\ and\ \citenamefont {Crispino}}]{Leite:2019eis}%
  \BibitemOpen
  \bibfield  {author} {\bibinfo {author} {\bibfnamefont {L.~C.~S.}\
  \bibnamefont {Leite}}, \bibinfo {author} {\bibfnamefont {C.~L.}\ \bibnamefont
  {Benone}}, \ and\ \bibinfo {author} {\bibfnamefont {L.~C.~B.}\ \bibnamefont
  {Crispino}},\ }\href {\doibase 10.1016/j.physletb.2019.06.027} {\bibfield
  {journal} {\bibinfo  {journal} {Phys. Lett.}\ }\textbf {\bibinfo {volume}
  {B795}},\ \bibinfo {pages} {496} (\bibinfo {year} {2019})},\ \Eprint
  {http://arxiv.org/abs/1907.04746} {arXiv:1907.04746 [gr-qc]} \BibitemShut
  {NoStop}%
\bibitem [{\citenamefont {Crispino}\ \emph {et~al.}(2015)\citenamefont
  {Crispino}, \citenamefont {Dolan}, \citenamefont {Higuchi},\ and\
  \citenamefont {de~Oliveira}}]{Crispino:2015gua}%
  \BibitemOpen
  \bibfield  {author} {\bibinfo {author} {\bibfnamefont {L.~C.~B.}\
  \bibnamefont {Crispino}}, \bibinfo {author} {\bibfnamefont {S.~R.}\
  \bibnamefont {Dolan}}, \bibinfo {author} {\bibfnamefont {A.}~\bibnamefont
  {Higuchi}}, \ and\ \bibinfo {author} {\bibfnamefont {E.~S.}\ \bibnamefont
  {de~Oliveira}},\ }\href {\doibase 10.1103/PhysRevD.92.084056} {\bibfield
  {journal} {\bibinfo  {journal} {Phys. Rev.}\ }\textbf {\bibinfo {volume}
  {D92}},\ \bibinfo {pages} {084056} (\bibinfo {year} {2015})},\ \Eprint
  {http://arxiv.org/abs/1507.03993} {arXiv:1507.03993 [gr-qc]} \BibitemShut
  {NoStop}%
\bibitem [{\citenamefont {Watson}(1918)}]{Watson18}%
  \BibitemOpen
  \bibfield  {author} {\bibinfo {author} {\bibfnamefont {G.~N.}\ \bibnamefont
  {Watson}},\ }\href@noop {} {\bibfield  {journal} {\bibinfo  {journal} {Proc.\
  R.\ Soc.\ London A}\ }\textbf {\bibinfo {volume} {95}},\ \bibinfo {pages}
  {83} (\bibinfo {year} {1918})}\BibitemShut {NoStop}%
\bibitem [{\citenamefont {Sommerfeld}(1949)}]{Sommerfeld49}%
  \BibitemOpen
  \bibfield  {author} {\bibinfo {author} {\bibfnamefont {A.}~\bibnamefont
  {Sommerfeld}},\ }\href@noop {} {\emph {\bibinfo {title} {Partial Differential
  Equations of Physics}}}\ (\bibinfo  {publisher} {Academic Press, New York},\
  \bibinfo {year} {1949})\BibitemShut {NoStop}%
\bibitem [{\citenamefont {de~Alfaro}\ and\ \citenamefont
  {Regge}(1965)}]{deAlfaro:1965zz}%
  \BibitemOpen
  \bibfield  {author} {\bibinfo {author} {\bibfnamefont {V.}~\bibnamefont
  {de~Alfaro}}\ and\ \bibinfo {author} {\bibfnamefont {T.}~\bibnamefont
  {Regge}},\ }\href@noop {} {\emph {\bibinfo {title} {{Potential
  Scattering}}}}\ (\bibinfo  {publisher} {North-Holland Publishing Company,
  Amsterdam},\ \bibinfo {year} {1965})\BibitemShut {NoStop}%
\bibitem [{\citenamefont {Newton}(1982)}]{Newton:1982qc}%
  \BibitemOpen
  \bibfield  {author} {\bibinfo {author} {\bibfnamefont {R.~G.}\ \bibnamefont
  {Newton}},\ }\href@noop {} {\emph {\bibinfo {title} {{Scattering Theory of
  Waves and Particles}}}},\ \bibinfo {edition} {2nd}\ ed.\ (\bibinfo
  {publisher} {Springer-Verlag, New York},\ \bibinfo {year} {1982})\BibitemShut
  {NoStop}%
\bibitem [{\citenamefont {Nussenzveig}(2006)}]{Nussenzveig:2006}%
  \BibitemOpen
  \bibfield  {author} {\bibinfo {author} {\bibfnamefont {H.~M.}\ \bibnamefont
  {Nussenzveig}},\ }\href@noop {} {\emph {\bibinfo {title} {Diffraction effects
  in semiclassical scattering}}},\ Vol.~\bibinfo {volume} {1}\ (\bibinfo
  {publisher} {Cambridge University Press},\ \bibinfo {year}
  {2006})\BibitemShut {NoStop}%
\bibitem [{\citenamefont {Grandy}(2000)}]{Grandy2000}%
  \BibitemOpen
  \bibfield  {author} {\bibinfo {author} {\bibfnamefont {W.~T.}\ \bibnamefont
  {Grandy}},\ }\href@noop {} {\emph {\bibinfo {title} {Scattering of Waves from
  Large Spheres}}}\ (\bibinfo  {publisher} {Cambridge University Press,
  Cambridge},\ \bibinfo {year} {2000})\BibitemShut {NoStop}%
\bibitem [{\citenamefont {{\"U}berall}(1992)}]{Uberall1992}%
  \BibitemOpen
  \bibfield  {author} {\bibinfo {author} {\bibfnamefont {H.}~\bibnamefont
  {{\"U}berall}},\ }\href@noop {} {\emph {\bibinfo {title} {Acoustic Resonance
  Scattering}}}\ (\bibinfo  {publisher} {Gordon and Breach, New York},\
  \bibinfo {year} {1992})\BibitemShut {NoStop}%
\bibitem [{\citenamefont {Aki}\ and\ \citenamefont
  {Richards}(2002)}]{AkiRichards2002}%
  \BibitemOpen
  \bibfield  {author} {\bibinfo {author} {\bibfnamefont {K.}~\bibnamefont
  {Aki}}\ and\ \bibinfo {author} {\bibfnamefont {P.}~\bibnamefont {Richards}},\
  }\href@noop {} {\emph {\bibinfo {title} {Quantitative Seismology}}},\
  \bibinfo {edition} {2nd}\ ed.\ (\bibinfo  {publisher} {University Science
  Book, Sausalito},\ \bibinfo {year} {2002})\BibitemShut {NoStop}%
\bibitem [{\citenamefont {Gribov}(2003)}]{Gribov69}%
  \BibitemOpen
  \bibfield  {author} {\bibinfo {author} {\bibfnamefont {V.~N.}\ \bibnamefont
  {Gribov}},\ }\href@noop {} {\emph {\bibinfo {title} {The Theory of Complex
  Angular Momenta: Gribov Lectures on Theoretical Physics}}}\ (\bibinfo
  {publisher} {Cambridge University Press, Cambridge},\ \bibinfo {year}
  {2003})\BibitemShut {NoStop}%
\bibitem [{\citenamefont {Collins}(1977)}]{Collins77}%
  \BibitemOpen
  \bibfield  {author} {\bibinfo {author} {\bibfnamefont {P.~D.~B.}\
  \bibnamefont {Collins}},\ }\href@noop {} {\emph {\bibinfo {title} {An
  Introduction to Regge Theory and High-Energy Physics}}}\ (\bibinfo
  {publisher} {Cambridge University Press, Cambridge},\ \bibinfo {year}
  {1977})\BibitemShut {NoStop}%
\bibitem [{\citenamefont {Barone}\ and\ \citenamefont
  {Predazzi}(2002)}]{BaronePredazzi2002}%
  \BibitemOpen
  \bibfield  {author} {\bibinfo {author} {\bibfnamefont {V.}~\bibnamefont
  {Barone}}\ and\ \bibinfo {author} {\bibfnamefont {E.}~\bibnamefont
  {Predazzi}},\ }\href@noop {} {\emph {\bibinfo {title} {High-Energy Particle
  Diffraction}}}\ (\bibinfo  {publisher} {Springer-Verlag, Berlin},\ \bibinfo
  {year} {2002})\BibitemShut {NoStop}%
\bibitem [{\citenamefont {Donnachie}\ \emph {et~al.}(2005)\citenamefont
  {Donnachie}, \citenamefont {Dosch}, \citenamefont {Landshoff},\ and\
  \citenamefont {Nachtmann}}]{DonnachieETAL2005}%
  \BibitemOpen
  \bibfield  {author} {\bibinfo {author} {\bibfnamefont {S.}~\bibnamefont
  {Donnachie}}, \bibinfo {author} {\bibfnamefont {G.}~\bibnamefont {Dosch}},
  \bibinfo {author} {\bibfnamefont {P.~V.}\ \bibnamefont {Landshoff}}, \ and\
  \bibinfo {author} {\bibfnamefont {O.}~\bibnamefont {Nachtmann}},\ }\href@noop
  {} {\emph {\bibinfo {title} {Pomeron Physics and QCD}}}\ (\bibinfo
  {publisher} {Cambridge University Press, Cambridge},\ \bibinfo {year}
  {2005})\BibitemShut {NoStop}%
\bibitem [{\citenamefont {Andersson}\ and\ \citenamefont
  {Thylwe}(1994)}]{Andersson:1994rk}%
  \BibitemOpen
  \bibfield  {author} {\bibinfo {author} {\bibfnamefont {N.}~\bibnamefont
  {Andersson}}\ and\ \bibinfo {author} {\bibfnamefont {K.~E.}\ \bibnamefont
  {Thylwe}},\ }\href {\doibase 10.1088/0264-9381/11/12/013} {\bibfield
  {journal} {\bibinfo  {journal} {Class. Quant. Grav.}\ }\textbf {\bibinfo
  {volume} {11}},\ \bibinfo {pages} {2991} (\bibinfo {year}
  {1994})}\BibitemShut {NoStop}%
\bibitem [{\citenamefont {Andersson}(1994)}]{Andersson:1994rm}%
  \BibitemOpen
  \bibfield  {author} {\bibinfo {author} {\bibfnamefont {N.}~\bibnamefont
  {Andersson}},\ }\href {\doibase 10.1088/0264-9381/11/12/014} {\bibfield
  {journal} {\bibinfo  {journal} {Class.\ Quant.\ Grav.}\ }\textbf {\bibinfo
  {volume} {11}},\ \bibinfo {pages} {3003} (\bibinfo {year}
  {1994})}\BibitemShut {NoStop}%
\bibitem [{\citenamefont {Decanini}\ \emph {et~al.}(2003)\citenamefont
  {Decanini}, \citenamefont {Folacci},\ and\ \citenamefont
  {Jensen}}]{Decanini:2002ha}%
  \BibitemOpen
  \bibfield  {author} {\bibinfo {author} {\bibfnamefont {Y.}~\bibnamefont
  {Decanini}}, \bibinfo {author} {\bibfnamefont {A.}~\bibnamefont {Folacci}}, \
  and\ \bibinfo {author} {\bibfnamefont {B.}~\bibnamefont {Jensen}},\ }\href
  {\doibase 10.1103/PhysRevD.67.124017} {\bibfield  {journal} {\bibinfo
  {journal} {Phys. Rev.}\ }\textbf {\bibinfo {volume} {D67}},\ \bibinfo {pages}
  {124017} (\bibinfo {year} {2003})},\ \Eprint
  {http://arxiv.org/abs/gr-qc/0212093} {arXiv:gr-qc/0212093 [gr-qc]}
  \BibitemShut {NoStop}%
\bibitem [{\citenamefont {Folacci}\ and\ \citenamefont {Ould
  El~Hadj}(2019{\natexlab{a}})}]{Folacci:2019cmc}%
  \BibitemOpen
  \bibfield  {author} {\bibinfo {author} {\bibfnamefont {A.}~\bibnamefont
  {Folacci}}\ and\ \bibinfo {author} {\bibfnamefont {M.}~\bibnamefont {Ould
  El~Hadj}},\ }\href {\doibase 10.1103/PhysRevD.99.104079} {\bibfield
  {journal} {\bibinfo  {journal} {Phys. Rev.}\ }\textbf {\bibinfo {volume}
  {D99}},\ \bibinfo {pages} {104079} (\bibinfo {year} {2019}{\natexlab{a}})},\
  \Eprint {http://arxiv.org/abs/1901.03965} {arXiv:1901.03965 [gr-qc]}
  \BibitemShut {NoStop}%
\bibitem [{\citenamefont {Folacci}\ and\ \citenamefont {Ould
  El~Hadj}(2019{\natexlab{b}})}]{Folacci:2019vtt}%
  \BibitemOpen
  \bibfield  {author} {\bibinfo {author} {\bibfnamefont {A.}~\bibnamefont
  {Folacci}}\ and\ \bibinfo {author} {\bibfnamefont {M.}~\bibnamefont {Ould
  El~Hadj}},\ }\href {\doibase 10.1103/PhysRevD.100.064009} {\bibfield
  {journal} {\bibinfo  {journal} {Phys. Rev.}\ }\textbf {\bibinfo {volume}
  {D100}},\ \bibinfo {pages} {064009} (\bibinfo {year} {2019}{\natexlab{b}})},\
  \Eprint {http://arxiv.org/abs/1906.01441} {arXiv:1906.01441 [gr-qc]}
  \BibitemShut {NoStop}%
\bibitem [{\citenamefont {Decanini}\ and\ \citenamefont
  {Folacci}(2010)}]{Decanini:2009mu}%
  \BibitemOpen
  \bibfield  {author} {\bibinfo {author} {\bibfnamefont {Y.}~\bibnamefont
  {Decanini}}\ and\ \bibinfo {author} {\bibfnamefont {A.}~\bibnamefont
  {Folacci}},\ }\href {\doibase 10.1103/PhysRevD.81.024031} {\bibfield
  {journal} {\bibinfo  {journal} {Phys.\ Rev.\ D}\ }\textbf {\bibinfo {volume}
  {81}},\ \bibinfo {pages} {024031} (\bibinfo {year} {2010})},\ \Eprint
  {http://arxiv.org/abs/0906.2601} {arXiv:0906.2601 [gr-qc]} \BibitemShut
  {NoStop}%
\bibitem [{\citenamefont {Decanini}\ \emph
  {et~al.}(2011{\natexlab{a}})\citenamefont {Decanini}, \citenamefont
  {Esposito-Farese},\ and\ \citenamefont {Folacci}}]{Decanini:2011xi}%
  \BibitemOpen
  \bibfield  {author} {\bibinfo {author} {\bibfnamefont {Y.}~\bibnamefont
  {Decanini}}, \bibinfo {author} {\bibfnamefont {G.}~\bibnamefont
  {Esposito-Farese}}, \ and\ \bibinfo {author} {\bibfnamefont {A.}~\bibnamefont
  {Folacci}},\ }\href {\doibase 10.1103/PhysRevD.83.044032} {\bibfield
  {journal} {\bibinfo  {journal} {Phys. Rev.}\ }\textbf {\bibinfo {volume}
  {D83}},\ \bibinfo {pages} {044032} (\bibinfo {year} {2011}{\natexlab{a}})},\
  \Eprint {http://arxiv.org/abs/1101.0781} {arXiv:1101.0781 [gr-qc]}
  \BibitemShut {NoStop}%
\bibitem [{\citenamefont {Decanini}\ \emph
  {et~al.}(2011{\natexlab{b}})\citenamefont {Decanini}, \citenamefont
  {Folacci},\ and\ \citenamefont {Raffaelli}}]{Decanini:2011xw}%
  \BibitemOpen
  \bibfield  {author} {\bibinfo {author} {\bibfnamefont {Y.}~\bibnamefont
  {Decanini}}, \bibinfo {author} {\bibfnamefont {A.}~\bibnamefont {Folacci}}, \
  and\ \bibinfo {author} {\bibfnamefont {B.}~\bibnamefont {Raffaelli}},\ }\href
  {\doibase 10.1088/0264-9381/28/17/175021} {\bibfield  {journal} {\bibinfo
  {journal} {Class. Quant. Grav.}\ }\textbf {\bibinfo {volume} {28}},\ \bibinfo
  {pages} {175021} (\bibinfo {year} {2011}{\natexlab{b}})},\ \Eprint
  {http://arxiv.org/abs/1104.3285} {arXiv:1104.3285 [gr-qc]} \BibitemShut
  {NoStop}%
\bibitem [{\citenamefont {Folacci}\ and\ \citenamefont {Ould
  El~Hadj}(2018)}]{Folacci:2018sef}%
  \BibitemOpen
  \bibfield  {author} {\bibinfo {author} {\bibfnamefont {A.}~\bibnamefont
  {Folacci}}\ and\ \bibinfo {author} {\bibfnamefont {M.}~\bibnamefont {Ould
  El~Hadj}},\ }\href {\doibase 10.1103/PhysRevD.98.064052} {\bibfield
  {journal} {\bibinfo  {journal} {Phys.\ Rev.\ D}\ }\textbf {\bibinfo {volume}
  {98}},\ \bibinfo {pages} {064052} (\bibinfo {year} {2018})},\ \Eprint
  {http://arxiv.org/abs/1807.09056} {arXiv:1807.09056 [gr-qc]} \BibitemShut
  {NoStop}%
\bibitem [{\citenamefont {Chandrasekhar}\ and\ \citenamefont
  {Ferrari}(1992)}]{ChandrasekharIV:1992ey}%
  \BibitemOpen
  \bibfield  {author} {\bibinfo {author} {\bibfnamefont {S.}~\bibnamefont
  {Chandrasekhar}}\ and\ \bibinfo {author} {\bibfnamefont {V.}~\bibnamefont
  {Ferrari}},\ }\href {\doibase 10.1098/rspa.1992.0051} {\bibfield  {journal}
  {\bibinfo  {journal} {Proc. Roy. Soc. Lond.}\ }\textbf {\bibinfo {volume}
  {A437}},\ \bibinfo {pages} {133} (\bibinfo {year} {1992})},\ \bibinfo {note}
  {[,1058(1992)]}\BibitemShut {NoStop}%
\bibitem [{\citenamefont {Giesler}\ \emph {et~al.}(2019)\citenamefont
  {Giesler}, \citenamefont {Isi}, \citenamefont {Scheel},\ and\ \citenamefont
  {Teukolsky}}]{Giesler:2019uxc}%
  \BibitemOpen
  \bibfield  {author} {\bibinfo {author} {\bibfnamefont {M.}~\bibnamefont
  {Giesler}}, \bibinfo {author} {\bibfnamefont {M.}~\bibnamefont {Isi}},
  \bibinfo {author} {\bibfnamefont {M.}~\bibnamefont {Scheel}}, \ and\ \bibinfo
  {author} {\bibfnamefont {S.}~\bibnamefont {Teukolsky}},\ }\href@noop {} {\
  (\bibinfo {year} {2019})},\ \Eprint {http://arxiv.org/abs/1903.08284}
  {arXiv:1903.08284 [gr-qc]} \BibitemShut {NoStop}%
\bibitem [{\citenamefont {Detweiler}\ and\ \citenamefont
  {Lindblom}(1985)}]{Detweiler:1985zz}%
  \BibitemOpen
  \bibfield  {author} {\bibinfo {author} {\bibfnamefont {S.~L.}\ \bibnamefont
  {Detweiler}}\ and\ \bibinfo {author} {\bibfnamefont {L.}~\bibnamefont
  {Lindblom}},\ }\href {\doibase 10.1086/163127} {\bibfield  {journal}
  {\bibinfo  {journal} {Astrophys. J.}\ }\textbf {\bibinfo {volume} {292}},\
  \bibinfo {pages} {12} (\bibinfo {year} {1985})}\BibitemShut {NoStop}%
\bibitem [{\citenamefont {Kokkotas}\ and\ \citenamefont
  {Schutz}(1986)}]{Kokkotas:1986gd}%
  \BibitemOpen
  \bibfield  {author} {\bibinfo {author} {\bibfnamefont {K.~D.}\ \bibnamefont
  {Kokkotas}}\ and\ \bibinfo {author} {\bibfnamefont {B.~F.}\ \bibnamefont
  {Schutz}},\ }\href@noop {} {\bibfield  {journal} {\bibinfo  {journal} {Gen.
  Rel. Grav.}\ }\textbf {\bibinfo {volume} {18}},\ \bibinfo {pages} {913}
  (\bibinfo {year} {1986})}\BibitemShut {NoStop}%
\bibitem [{\citenamefont {Chandrasekhar}\ and\ \citenamefont
  {Ferrari}(1991)}]{Chandrasekhar449}%
  \BibitemOpen
  \bibfield  {author} {\bibinfo {author} {\bibfnamefont {S.}~\bibnamefont
  {Chandrasekhar}}\ and\ \bibinfo {author} {\bibfnamefont {V.}~\bibnamefont
  {Ferrari}},\ }\href {\doibase 10.1098/rspa.1991.0104} {\bibfield  {journal}
  {\bibinfo  {journal} {Proceedings of the Royal Society of London A:
  Mathematical, Physical and Engineering Sciences}\ }\textbf {\bibinfo {volume}
  {434}},\ \bibinfo {pages} {449} (\bibinfo {year} {1991})}\BibitemShut
  {NoStop}%
\bibitem [{\citenamefont {Kokkotas}\ and\ \citenamefont
  {Schutz}(1992)}]{Kokkotas:1992ka}%
  \BibitemOpen
  \bibfield  {author} {\bibinfo {author} {\bibfnamefont {K.~D.}\ \bibnamefont
  {Kokkotas}}\ and\ \bibinfo {author} {\bibfnamefont {B.~F.}\ \bibnamefont
  {Schutz}},\ }\href@noop {} {\bibfield  {journal} {\bibinfo  {journal} {Mon.
  Not. Roy. Astron. Soc.}\ }\textbf {\bibinfo {volume} {225}},\ \bibinfo
  {pages} {119} (\bibinfo {year} {1992})}\BibitemShut {NoStop}%
\bibitem [{\citenamefont {Leins}\ \emph {et~al.}(1993)\citenamefont {Leins},
  \citenamefont {Nollert},\ and\ \citenamefont {Soffel}}]{Leins:1993zz}%
  \BibitemOpen
  \bibfield  {author} {\bibinfo {author} {\bibfnamefont {M.}~\bibnamefont
  {Leins}}, \bibinfo {author} {\bibfnamefont {H.~P.}\ \bibnamefont {Nollert}},
  \ and\ \bibinfo {author} {\bibfnamefont {M.~H.}\ \bibnamefont {Soffel}},\
  }\href {\doibase 10.1103/PhysRevD.48.3467} {\bibfield  {journal} {\bibinfo
  {journal} {Phys. Rev.}\ }\textbf {\bibinfo {volume} {D48}},\ \bibinfo {pages}
  {3467} (\bibinfo {year} {1993})}\BibitemShut {NoStop}%
\bibitem [{\citenamefont {Andersson}\ \emph {et~al.}(1996)\citenamefont
  {Andersson}, \citenamefont {Kojima},\ and\ \citenamefont
  {Kokkotas}}]{Andersson:1995ez}%
  \BibitemOpen
  \bibfield  {author} {\bibinfo {author} {\bibfnamefont {N.}~\bibnamefont
  {Andersson}}, \bibinfo {author} {\bibfnamefont {Y.}~\bibnamefont {Kojima}}, \
  and\ \bibinfo {author} {\bibfnamefont {K.~D.}\ \bibnamefont {Kokkotas}},\
  }\href {\doibase 10.1086/177199} {\bibfield  {journal} {\bibinfo  {journal}
  {Astrophys. J.}\ }\textbf {\bibinfo {volume} {462}},\ \bibinfo {pages} {855}
  (\bibinfo {year} {1996})},\ \Eprint {http://arxiv.org/abs/gr-qc/9512048}
  {arXiv:gr-qc/9512048 [gr-qc]} \BibitemShut {NoStop}%
\bibitem [{\citenamefont {Andersson}(1996)}]{Andersson1996}%
  \BibitemOpen
  \bibfield  {author} {\bibinfo {author} {\bibfnamefont {N.}~\bibnamefont
  {Andersson}},\ }\href {\doibase 10.1007/BF02113773} {\bibfield  {journal}
  {\bibinfo  {journal} {General Relativity and Gravitation}\ }\textbf {\bibinfo
  {volume} {28}},\ \bibinfo {pages} {1433} (\bibinfo {year}
  {1996})}\BibitemShut {NoStop}%
\bibitem [{\citenamefont {Kokkotas}\ and\ \citenamefont
  {Schmidt}(1999)}]{Kokkotas:1999bd}%
  \BibitemOpen
  \bibfield  {author} {\bibinfo {author} {\bibfnamefont {K.~D.}\ \bibnamefont
  {Kokkotas}}\ and\ \bibinfo {author} {\bibfnamefont {B.~G.}\ \bibnamefont
  {Schmidt}},\ }\href {\doibase 10.12942/lrr-1999-2} {\bibfield  {journal}
  {\bibinfo  {journal} {Living Rev. Rel.}\ }\textbf {\bibinfo {volume} {2}},\
  \bibinfo {pages} {2} (\bibinfo {year} {1999})},\ \Eprint
  {http://arxiv.org/abs/gr-qc/9909058} {arXiv:gr-qc/9909058 [gr-qc]}
  \BibitemShut {NoStop}%
\bibitem [{\citenamefont {Dolan}\ and\ \citenamefont
  {Stratton}(2017)}]{Dolan:2017rtj}%
  \BibitemOpen
  \bibfield  {author} {\bibinfo {author} {\bibfnamefont {S.~R.}\ \bibnamefont
  {Dolan}}\ and\ \bibinfo {author} {\bibfnamefont {T.}~\bibnamefont
  {Stratton}},\ }\href {\doibase 10.1103/PhysRevD.95.124055} {\bibfield
  {journal} {\bibinfo  {journal} {Phys. Rev.}\ }\textbf {\bibinfo {volume}
  {D95}},\ \bibinfo {pages} {124055} (\bibinfo {year} {2017})},\ \Eprint
  {http://arxiv.org/abs/1702.06127} {arXiv:1702.06127 [gr-qc]} \BibitemShut
  {NoStop}%
\bibitem [{\citenamefont {Stratton}\ and\ \citenamefont
  {Dolan}(2019)}]{Stratton:2019deq}%
  \BibitemOpen
  \bibfield  {author} {\bibinfo {author} {\bibfnamefont {T.}~\bibnamefont
  {Stratton}}\ and\ \bibinfo {author} {\bibfnamefont {S.~R.}\ \bibnamefont
  {Dolan}},\ }\href {\doibase 10.1103/PhysRevD.100.024007} {\bibfield
  {journal} {\bibinfo  {journal} {Phys. Rev.}\ }\textbf {\bibinfo {volume}
  {D100}},\ \bibinfo {pages} {024007} (\bibinfo {year} {2019})},\ \Eprint
  {http://arxiv.org/abs/1903.00025} {arXiv:1903.00025 [gr-qc]} \BibitemShut
  {NoStop}%
\bibitem [{Note1()}]{Note1}%
  \BibitemOpen
  \bibinfo {note} {The rainbow angle $\theta _r \approx 4M/R$ is distinct from
  the Einstein ring angle of $\theta _E \approx \protect \sqrt
  {4M/r}$}\BibitemShut {NoStop}%
\bibitem [{\citenamefont {Nambu}\ \emph {et~al.}(2019)\citenamefont {Nambu},
  \citenamefont {Noda},\ and\ \citenamefont {Sakai}}]{Nambu:2019sqn}%
  \BibitemOpen
  \bibfield  {author} {\bibinfo {author} {\bibfnamefont {Y.}~\bibnamefont
  {Nambu}}, \bibinfo {author} {\bibfnamefont {S.}~\bibnamefont {Noda}}, \ and\
  \bibinfo {author} {\bibfnamefont {Y.}~\bibnamefont {Sakai}},\ }\href@noop {}
  {\  (\bibinfo {year} {2019})},\ \Eprint {http://arxiv.org/abs/1905.01793}
  {arXiv:1905.01793 [gr-qc]} \BibitemShut {NoStop}%
\bibitem [{\citenamefont {He}(2019)}]{He:2019orl}%
  \BibitemOpen
  \bibfield  {author} {\bibinfo {author} {\bibfnamefont {J.-h.}\ \bibnamefont
  {He}},\ }\href@noop {} {\  (\bibinfo {year} {2019})},\ \Eprint
  {http://arxiv.org/abs/1912.00325} {arXiv:1912.00325 [gr-qc]} \BibitemShut
  {NoStop}%
\bibitem [{\citenamefont {Alexandre}\ and\ \citenamefont
  {Clough}(2018)}]{Alexandre:2018crg}%
  \BibitemOpen
  \bibfield  {author} {\bibinfo {author} {\bibfnamefont {J.}~\bibnamefont
  {Alexandre}}\ and\ \bibinfo {author} {\bibfnamefont {K.}~\bibnamefont
  {Clough}},\ }\href {\doibase 10.1103/PhysRevD.98.043004} {\bibfield
  {journal} {\bibinfo  {journal} {Phys. Rev.}\ }\textbf {\bibinfo {volume}
  {D98}},\ \bibinfo {pages} {043004} (\bibinfo {year} {2018})},\ \Eprint
  {http://arxiv.org/abs/1805.01874} {arXiv:1805.01874 [hep-ph]} \BibitemShut
  {NoStop}%
\bibitem [{\citenamefont {Marchant}\ \emph {et~al.}(2019)\citenamefont
  {Marchant}, \citenamefont {Breivik}, \citenamefont {Berry}, \citenamefont
  {Mandel},\ and\ \citenamefont {Larson}}]{Marchant:2019swq}%
  \BibitemOpen
  \bibfield  {author} {\bibinfo {author} {\bibfnamefont {P.}~\bibnamefont
  {Marchant}}, \bibinfo {author} {\bibfnamefont {K.}~\bibnamefont {Breivik}},
  \bibinfo {author} {\bibfnamefont {C.~P.~L.}\ \bibnamefont {Berry}}, \bibinfo
  {author} {\bibfnamefont {I.}~\bibnamefont {Mandel}}, \ and\ \bibinfo {author}
  {\bibfnamefont {S.~L.}\ \bibnamefont {Larson}},\ }\href@noop {} {\  (\bibinfo
  {year} {2019})},\ \Eprint {http://arxiv.org/abs/1912.04268} {arXiv:1912.04268
  [astro-ph.SR]} \BibitemShut {NoStop}%
\bibitem [{\citenamefont {Voje~Johansen}\ and\ \citenamefont
  {Ravndal}(2006)}]{VojeJohansen:2005nd}%
  \BibitemOpen
  \bibfield  {author} {\bibinfo {author} {\bibfnamefont {N.}~\bibnamefont
  {Voje~Johansen}}\ and\ \bibinfo {author} {\bibfnamefont {F.}~\bibnamefont
  {Ravndal}},\ }\href {\doibase 10.1007/s10714-006-0242-0} {\bibfield
  {journal} {\bibinfo  {journal} {Gen. Rel. Grav.}\ }\textbf {\bibinfo {volume}
  {38}},\ \bibinfo {pages} {537} (\bibinfo {year} {2006})},\ \Eprint
  {http://arxiv.org/abs/physics/0508163} {arXiv:physics/0508163 [physics]}
  \BibitemShut {NoStop}%
\bibitem [{\citenamefont {Shapiro}\ and\ \citenamefont
  {Teukolsky}(1983)}]{Shapiro1983}%
  \BibitemOpen
  \bibfield  {author} {\bibinfo {author} {\bibfnamefont {S.~L.}\ \bibnamefont
  {Shapiro}}\ and\ \bibinfo {author} {\bibfnamefont {S.~A.}\ \bibnamefont
  {Teukolsky}},\ }\href@noop {} {\emph {\bibinfo {title} {Black Holes, White
  Dwarfs, and Neutron Stars: The Physics of Compact Objects}}}\ (\bibinfo
  {publisher} {Wiley, New-York},\ \bibinfo {year} {1983})\BibitemShut {NoStop}%
\bibitem [{\citenamefont {Cardoso}\ \emph {et~al.}(2014)\citenamefont
  {Cardoso}, \citenamefont {Crispino}, \citenamefont {Macedo}, \citenamefont
  {Okawa},\ and\ \citenamefont {Pani}}]{Cardoso:2014sna}%
  \BibitemOpen
  \bibfield  {author} {\bibinfo {author} {\bibfnamefont {V.}~\bibnamefont
  {Cardoso}}, \bibinfo {author} {\bibfnamefont {L.~C.~B.}\ \bibnamefont
  {Crispino}}, \bibinfo {author} {\bibfnamefont {C.~F.~B.}\ \bibnamefont
  {Macedo}}, \bibinfo {author} {\bibfnamefont {H.}~\bibnamefont {Okawa}}, \
  and\ \bibinfo {author} {\bibfnamefont {P.}~\bibnamefont {Pani}},\ }\href
  {\doibase 10.1103/PhysRevD.90.044069} {\bibfield  {journal} {\bibinfo
  {journal} {Phys. Rev. D}\ }\textbf {\bibinfo {volume} {D90}},\ \bibinfo
  {pages} {044069} (\bibinfo {year} {2014})},\ \Eprint
  {http://arxiv.org/abs/1406.5510} {arXiv:1406.5510 [gr-qc]} \BibitemShut
  {NoStop}%
\bibitem [{\citenamefont {Macedo}\ \emph {et~al.}(2018)\citenamefont {Macedo},
  \citenamefont {Stratton}, \citenamefont {Dolan},\ and\ \citenamefont
  {Crispino}}]{Macedo:2018yoi}%
  \BibitemOpen
  \bibfield  {author} {\bibinfo {author} {\bibfnamefont {C.~F.~B.}\
  \bibnamefont {Macedo}}, \bibinfo {author} {\bibfnamefont {T.}~\bibnamefont
  {Stratton}}, \bibinfo {author} {\bibfnamefont {S.}~\bibnamefont {Dolan}}, \
  and\ \bibinfo {author} {\bibfnamefont {L.~C.~B.}\ \bibnamefont {Crispino}},\
  }\href {\doibase 10.1103/PhysRevD.98.104034} {\bibfield  {journal} {\bibinfo
  {journal} {Phys. Rev.}\ }\textbf {\bibinfo {volume} {D98}},\ \bibinfo {pages}
  {104034} (\bibinfo {year} {2018})},\ \Eprint
  {http://arxiv.org/abs/1807.04762} {arXiv:1807.04762 [gr-qc]} \BibitemShut
  {NoStop}%
\bibitem [{\citenamefont {Andersson}\ and\ \citenamefont
  {Kokkotas}(1998)}]{Andersson:1997eq}%
  \BibitemOpen
  \bibfield  {author} {\bibinfo {author} {\bibfnamefont {N.}~\bibnamefont
  {Andersson}}\ and\ \bibinfo {author} {\bibfnamefont {K.~D.}\ \bibnamefont
  {Kokkotas}},\ }\href {\doibase 10.1046/j.1365-8711.1998.01541.x} {\bibfield
  {journal} {\bibinfo  {journal} {Mon. Not. Roy. Astron. Soc.}\ }\textbf
  {\bibinfo {volume} {297}},\ \bibinfo {pages} {493} (\bibinfo {year}
  {1998})},\ \Eprint {http://arxiv.org/abs/gr-qc/9706010} {arXiv:gr-qc/9706010
  [gr-qc]} \BibitemShut {NoStop}%
\bibitem [{\citenamefont {Thorne}\ and\ \citenamefont
  {Campolattaro}(1967)}]{thorne1967non}%
  \BibitemOpen
  \bibfield  {author} {\bibinfo {author} {\bibfnamefont {K.~S.}\ \bibnamefont
  {Thorne}}\ and\ \bibinfo {author} {\bibfnamefont {A.}~\bibnamefont
  {Campolattaro}},\ }\href@noop {} {\bibfield  {journal} {\bibinfo  {journal}
  {The Astrophysical Journal}\ }\textbf {\bibinfo {volume} {149}},\ \bibinfo
  {pages} {591} (\bibinfo {year} {1967})}\BibitemShut {NoStop}%
\bibitem [{\citenamefont {Leaver}(1990)}]{Leaver:1990zz}%
  \BibitemOpen
  \bibfield  {author} {\bibinfo {author} {\bibfnamefont {E.~W.}\ \bibnamefont
  {Leaver}},\ }\href {\doibase 10.1103/PhysRevD.41.2986} {\bibfield  {journal}
  {\bibinfo  {journal} {Phys. Rev.}\ }\textbf {\bibinfo {volume} {D41}},\
  \bibinfo {pages} {2986} (\bibinfo {year} {1990})}\BibitemShut {NoStop}%
\bibitem [{\citenamefont {Majumdar}\ and\ \citenamefont
  {Panchapakesan}(1989)}]{mp}%
  \BibitemOpen
  \bibfield  {author} {\bibinfo {author} {\bibfnamefont {B.}~\bibnamefont
  {Majumdar}}\ and\ \bibinfo {author} {\bibfnamefont {N.}~\bibnamefont
  {Panchapakesan}},\ }\href {\doibase 10.1103/PhysRevD.40.2568} {\bibfield
  {journal} {\bibinfo  {journal} {Phys.\ Rev.\ D}\ }\textbf {\bibinfo {volume}
  {40}},\ \bibinfo {pages} {2568} (\bibinfo {year} {1989})}\BibitemShut
  {NoStop}%
\bibitem [{\citenamefont {Leaver}(1985)}]{Leaver:1985ax}%
  \BibitemOpen
  \bibfield  {author} {\bibinfo {author} {\bibfnamefont {E.~W.}\ \bibnamefont
  {Leaver}},\ }\href {\doibase 10.1098/rspa.1985.0119} {\bibfield  {journal}
  {\bibinfo  {journal} {Proc.\ Roy.\ Soc.\ Lond.\ A}\ }\textbf {\bibinfo
  {volume} {402}},\ \bibinfo {pages} {285} (\bibinfo {year}
  {1985})}\BibitemShut {NoStop}%
\bibitem [{\citenamefont {Zhang}\ \emph {et~al.}(2011)\citenamefont {Zhang},
  \citenamefont {Wu},\ and\ \citenamefont {Leung}}]{Zhang:2011pq}%
  \BibitemOpen
  \bibfield  {author} {\bibinfo {author} {\bibfnamefont {Y.~J.}\ \bibnamefont
  {Zhang}}, \bibinfo {author} {\bibfnamefont {J.}~\bibnamefont {Wu}}, \ and\
  \bibinfo {author} {\bibfnamefont {P.~T.}\ \bibnamefont {Leung}},\ }\href
  {\doibase 10.1103/PhysRevD.83.064012} {\bibfield  {journal} {\bibinfo
  {journal} {Phys. Rev.}\ }\textbf {\bibinfo {volume} {D83}},\ \bibinfo {pages}
  {064012} (\bibinfo {year} {2011})},\ \Eprint {http://arxiv.org/abs/1101.0319}
  {arXiv:1101.0319 [gr-qc]} \BibitemShut {NoStop}%
\bibitem [{\citenamefont {Völkel}\ and\ \citenamefont
  {Kokkotas}(2019)}]{Volkel:2019gpq}%
  \BibitemOpen
  \bibfield  {author} {\bibinfo {author} {\bibfnamefont {S.~H.}\ \bibnamefont
  {Völkel}}\ and\ \bibinfo {author} {\bibfnamefont {K.~D.}\ \bibnamefont
  {Kokkotas}},\ }\href {\doibase 10.1088/1361-6382/ab186e} {\bibfield
  {journal} {\bibinfo  {journal} {Class. Quant. Grav.}\ }\textbf {\bibinfo
  {volume} {36}},\ \bibinfo {pages} {115002} (\bibinfo {year} {2019})},\
  \Eprint {http://arxiv.org/abs/1901.11262} {arXiv:1901.11262 [gr-qc]}
  \BibitemShut {NoStop}%
\bibitem [{\citenamefont {Berry}(1982)}]{Berry_1982}%
  \BibitemOpen
  \bibfield  {author} {\bibinfo {author} {\bibfnamefont {M.~V.}\ \bibnamefont
  {Berry}},\ }\href {\doibase 10.1088/0305-4470/15/12/021} {\bibfield
  {journal} {\bibinfo  {journal} {Journal of Physics A: Mathematical and
  General}\ }\textbf {\bibinfo {volume} {15}},\ \bibinfo {pages} {3693}
  (\bibinfo {year} {1982})}\BibitemShut {NoStop}%
\bibitem [{\citenamefont {Olver}\ \emph {et~al.}(2010)\citenamefont {Olver},
  \citenamefont {Lozier}, \citenamefont {Boisvert},\ and\ \citenamefont
  {Clark}}]{nist}%
  \BibitemOpen
  \bibfield  {author} {\bibinfo {author} {\bibfnamefont {F.~W.~J.}\
  \bibnamefont {Olver}}, \bibinfo {author} {\bibfnamefont {D.~W.}\ \bibnamefont
  {Lozier}}, \bibinfo {author} {\bibfnamefont {R.~F.}\ \bibnamefont
  {Boisvert}}, \ and\ \bibinfo {author} {\bibfnamefont {C.~W.}\ \bibnamefont
  {Clark}},\ }\href@noop {} {\emph {\bibinfo {title} {{NIST Handbook of
  Mathematical Functions}}}}\ (\bibinfo  {publisher} {Cambridge University
  Press},\ \bibinfo {year} {2010})\BibitemShut {NoStop}%
\bibitem [{\citenamefont {Abramowitz}\ and\ \citenamefont
  {Stegun}(1965)}]{AS65}%
  \BibitemOpen
  \bibfield  {author} {\bibinfo {author} {\bibfnamefont {M.}~\bibnamefont
  {Abramowitz}}\ and\ \bibinfo {author} {\bibfnamefont {I.~A.}\ \bibnamefont
  {Stegun}},\ }\href@noop {} {\emph {\bibinfo {title} {Handbook of Mathematical
  Functions}}}\ (\bibinfo  {publisher} {Dover, New-York},\ \bibinfo {year}
  {1965})\BibitemShut {NoStop}%
\bibitem [{\citenamefont {Bottino}\ and\ \citenamefont
  {Longoni}(1962)}]{Bottino1962}%
  \BibitemOpen
  \bibfield  {author} {\bibinfo {author} {\bibfnamefont {A.}~\bibnamefont
  {Bottino}}\ and\ \bibinfo {author} {\bibfnamefont {A.~M.}\ \bibnamefont
  {Longoni}},\ }\href {\doibase 10.1007/BF02745656} {\bibfield  {journal}
  {\bibinfo  {journal} {Il Nuovo Cimento (1955-1965)}\ }\textbf {\bibinfo
  {volume} {24}},\ \bibinfo {pages} {353} (\bibinfo {year} {1962})}\BibitemShut
  {NoStop}%
\end{thebibliography}%
\bibliographystyle{apsrev4-1}

\end{document}